\documentclass[fleqn,usenatbib,useAMS]{mnras}

\usepackage{graphicx}	
\usepackage{amsmath}	
\usepackage{amssymb}	
\usepackage{multicol}        
\usepackage{pdflscape}	
\usepackage{cancel}     
\usepackage{subcaption}
\usepackage{caption}
\usepackage{orcidlink}
\setcitestyle{notesep={ }}

\newcommand{\ahod}{\textsc{AbacusHOD}}
\newcommand{\sandy}[1]{\textcolor{black}{#1}}

\usepackage[T1]{fontenc}
\usepackage{ae,aecompl}

\usepackage{newtxtext,newtxmath}

\title[$k$NN-emulator]{Robust cosmological inference from non-linear scales with $k$-th nearest neighbor statistics}

\author[Yuan et al]{
Sihan Yuan,\thanks{E-mail: sihany@stanford.edu}\orcidlink{0000-0002-5992-7586}
Tom Abel,
and Risa H. Wechsler\orcidlink{0000-0003-2229-011X}
\\
Kavli Institute for Particle Astrophysics and Cosmology, Stanford University, 452 Lomita Mall, Stanford, CA 94305, USA\\
Department of Physics, Stanford University, 382 Via Pueblo Mall, Stanford, CA 94305, USA\\
SLAC National Accelerator Laboratory, 2575 Sand Hill Road, Menlo Park, CA  94025, USA
}

\date{Accepted XXX. Received YYY; in original form ZZZ}

\pubyear{2022}

\begin{document}
\label{firstpage}
\pagerange{\pageref{firstpage}--\pageref{lastpage}}
\maketitle

\begin{abstract}
We present the methodology for deriving accurate and reliable cosmological constraints from non-linear scales ($<50h^{-1}$Mpc) with $k$-th nearest neighbor ($k$NN) statistics. We detail our methods for choosing robust minimum scale cuts and validating galaxy--halo connection models. Using cross-validation, we identify the galaxy–halo model that ensures both good fits and unbiased predictions across diverse summary statistics. We demonstrate that we can model $k$NNs effectively down to transverse scales of $r_p\sim 3h^{-1}$Mpc and achieve precise and unbiased constraints on the matter density and clustering amplitude, leading to a 2\% constraint on $\sigma_8$. Our simulation-based model pipeline is resilient to varied model systematics, spanning simulation codes, halo finding, and cosmology priors. We demonstrate the effectiveness of this approach through an application to the Beyond-2p mock challenge. We propose further explorations to test more complex galaxy–halo connection models and tackle potential observational systematics. 
\end{abstract}





\section{Introduction}

The spatial distribution of galaxies presents one of the most powerful probes of the fundamental properties of the universe. A new generation of wide-area spectroscopic surveys, such as the Dark Energy Spectroscopic Instrument \citep[DESI;][]{2013Levi}, the Subaru Prime Focus Spectrograph \citep[PFS;][]{2014Takada}, the ESA \textit{Euclid} satellite mission \citep[][]{2011Laureijs}, and the NASA \textit{Roman Space Telescope} \citep[WFIRST;][]{2013Spergel} will enable us to study the 3D large-scale structure (LSS) of the universe with unprecedented precision. 
This precision promises insights into critical questions, including the cause of the universe's accelerated expansion and the theory of gravity, the nature of dark matter and neutrinos, the physics of the primordial universe, and the detailed process of galaxy formation.

Over recent decades, LSS analyses have seen significant advancements. On large scales ($> 50\,h^{-1}\,$Mpc), concise analytic models based on perturbation theory (of the density contrast) are sufficiently accurate. Given that density field on large scales is well approximated by a Gaussian random field, the 2-point correlation function (2PCF, or its Fourier pair the power spectrum) captures its complete information content. Consequently, large-scale 2PCF analyses using perturbative methods have become pivotal in modern cosmology. However, there is a wealth of additional information on smaller, non-linear scales.  These scales contain complex features in the density field and offer the highest signal-to-noise clustering measurements, but pose significant analytical challenges due to the intermingling of cosmology and galaxy physics.

Addressing these challenges and harnessing the full potential of small-scale information necessitate alternative modeling frameworks. While gravity-only $N$-body simulations can predict matter density accurately to smaller scales, the sheer computational expense and the precision requirements of modern surveys mean we need large volume, high-resolution simulations. To bridge this gap, surrogate models or emulators interpolate between a limited set of simulated cosmologies, providing efficient predictors for arbitrary cosmologies \citep[e.g.][]{2019DeRose, 2019Nishimichi, 2021Maksimova}. This method, termed simulation-based modeling, has been successfully implemented in various recent studies \citep[e.g.][]{2021Lange, 2022bYuan, 2021Kobayashi, 2021Chapman, 2022Zhai}.

Still, the credibility of these models on smaller scales remains a concern. Large-scale perturbative models use a sequence of bias parameters, keeping them physics-agnostic, to marginalise over the connection between galaxies and dark matter. On smaller scales, however, simulation-based methods require well-motivated models linking galaxies to dark matter haloes. The most commonly used approach is Halo Occupation Distribution \citep[HOD;][]{2000Peacock, 2001Scoccimarro, 2002Berlind, 2003Berlind, 2005Zheng, 2007bZheng} models due to its speed and flexibility.  In its simplest form, only halo mass is used, and the HOD can be summarised as $P(N_\mathrm{galaxy}|M_\mathrm{halo})$; a commonly used form of this model is summarised by \cite{2007bZheng} (also see section~\ref{subsec:model}). More recent studies using additional data and summary statistics have found the need to include a secondary halo property \citep[e.g.][]{2021bYuan, 2022Wang, 2022Beltz-Mohrmann, 2023Contreras}, an effect known as galaxy assembly bias \citep{2005Gao,2006Wechsler}. Although the HOD model has undergone extensive refinement, its inherent empirical nature and simplicity indicates that it may fail in detail when confronted with ever more precise data. Thus, a pivotal question is ensuring that the model connecting connect galaxies to dark matter is sufficiently flexible to avoid systematic biases in the cosmology inference. 

To date, most simulation-based studies have adopted a fiducial model for the galaxy--halo connection without conducting extensive tests to demonstrate its robustness. Some studies have shown convergence between a few different HOD models \citep{2022bYuan}, or for a wider range of models for the galaxy--halo connection \citep{2014Reddick}, but they have not typically demonstrated that such models span the necessary model space. This is essential in the era of precision cosmology given that it has been shown that different galaxy assembly bias assumptions can impact cosmological parameters \citep{2019Lange}. Here, we present new tests that can significantly increase our confidence in the assumed galaxy model. 

Beyond the question of robustness, modeling non-linear scales requires advanced summary statistics. While the large-scale density field is adequately represented by the 2PCF, non-linear scales require a more complex approach. The $k$-th nearest neighbor statistics \citep[$k$NN][]{2021Banerjee, 2022Banerjee, 2023Yuan} have emerged as a promising solution due to their informative nature, computational ease, and interpretative clarity.


In this work, we describe in detail the methodology we use to derive precise yet dependable cosmology constraints with $k$NNs from non-linear scales in the Beyond-2p blind mock challenge \citep{b2p}. The paper is structured as follows. Section~\ref{sec:data} describes the blind mock data. Section~\ref{sec:method} details our simulation-based modeling framework, including simulations, the HOD, our summary statistics, and emulation. Our results, accompanied by methods for selecting scale cuts and validating HOD modeling, are presented in section~\ref{sec:results}. We discuss our methodology's limitations and forecast potential advancements in section~\ref{sec:discuss}. Conclusions are drawn in section~\ref{sec:conclude}.

\sandy{Throughout this paper, we adopt the Planck 2018 $\Lambda$CDM cosmology ($\Omega_c h^2 = 0.1200$, $\Omega_b h^2 = 0.02237$, $\sigma_8 = 0.811355$, $n_s = 0.9649$, $h = 0.6736$, $w_0 = -1$, and $w_a = 0$) \citep{2020Planck}.}

\section{Blind mock data}
\label{sec:data}

To test our cosmology inference framework, we use the blind mock galaxy catalogs described in \cite{b2p}. \sandy{The significance of this mock lies in that the true cosmology deviates significantly from Planck and is kept secret until all analyses were done and finalised. The galaxy model is also kept secret and will be hidden in the future to enable future participation in the blind challenge. In context, this is the first blind mock challenge specifically designed for analyses pipelines that target non-linear scales, with sufficient volume to match the new generation of cosmology surveys.}

The mock catalogs are created by populating galaxies in $N$-body simulation halo catalogs at $z = 1$, following an HOD prescription designed by the creators of the Beyond-2p challenge. 
While the mocks assume a flat $\Lambda$CDM cosmology in which the CMB acoustic scale $\theta_\star$ is fixed to the value adopted in \textsc{AbacusSummit} \sandy{(see section~\ref{sec:abacus})}, the input values of other cosmological parameters and the HOD model are kept blind.
For creating the halo catalogs, dark-matter-only simulations are run in $(2\,h^{-1}\,\mathrm{Gpc})^3$ boxes using the \textsc{GINKAKU} $N$-body solver \citep{ginkaku}.
The initial conditions are generated using the second-order Lagrangian perturbation theory \citep{Crocce:2006ve} at $z = 49$, and haloes are identified using the \textsc{Rockstar} halo finder \citep{2013Behroozi}.
The redshift-space distortions are implemented by modulating each galaxy's position along the $z$-axis with the redshift-space displacement determined by its velocity. \sandy{An additional ten $(2\,h^{-1}\,\mathrm{Gpc})^3$ boxes are generated with the same cosmology and HOD, but different initial phases for covariance calculations. }

\sandy{It is important to point out that these mocks are tuned to roughly produce the correct spatial number density and linear bias of DESI Luminous Red Galaxies, but we were not given any additional details about the underlying model beyond the fact that it is an HOD. We do not know the form of the HOD, the value of any HOD parameters, or the inclusion of any non-vanilla extensions such as galaxy assembly bias. The fact that the HOD is implemented on top of a new simulation and halo code means that there will be significant and complex differences between the mock HOD and any existing implementations. A key aim of this paper is to quantify to what extent we can marginalise over these differences and uncertainties about the model. }

\section{Methodology}
\label{sec:method}
We use a simulation-based model for this analysis. In this section, we introduce the various layers of the model, including the simulation suite, the HOD model, the $k$NN and 2PCF summary statistics, and our neural-net-based emulator. 
\subsection{\textsc{AbacusSummit} simulations}
\label{sec:abacus}

The \textsc{AbacusSummit} simulation suite \citep[][]{2021Maksimova} is a set of large, high-accuracy cosmological N-body simulations using the \textsc{Abacus} N-body code \citep{2019Garrison, 2021bGarrison}, designed to exceed the cosmological simulation requirements of the Dark Energy Spectroscopic Instrument (DESI) survey \citep{2013Levi}. \textsc{AbacusSummit} consists of over 150 simulations, containing approximately 60 trillion particles at 97 different cosmologies. 
For this analysis, we use exclusively the ``base'' configuration boxes in the simulation suite, each of which contains $6912^3$ particles within a $(2h^{-1}$Gpc$)^3$ volume, corresponding to a particle mass of $2.1 \times 10^9 h^{-1}M_\odot$. \footnote{For more details, see \url{https://abacussummit.readthedocs.io/en/latest/abacussummit.html}}
The \textsc{AbacusSummit} suite also uses a specialised spherical-overdensity based halo finder known as {\sc CompaSO} \citep{2021Hadzhiyska}.

\begin{figure}
    \centering
    \hspace*{-0.3cm}
    \includegraphics[width = 3.6in]{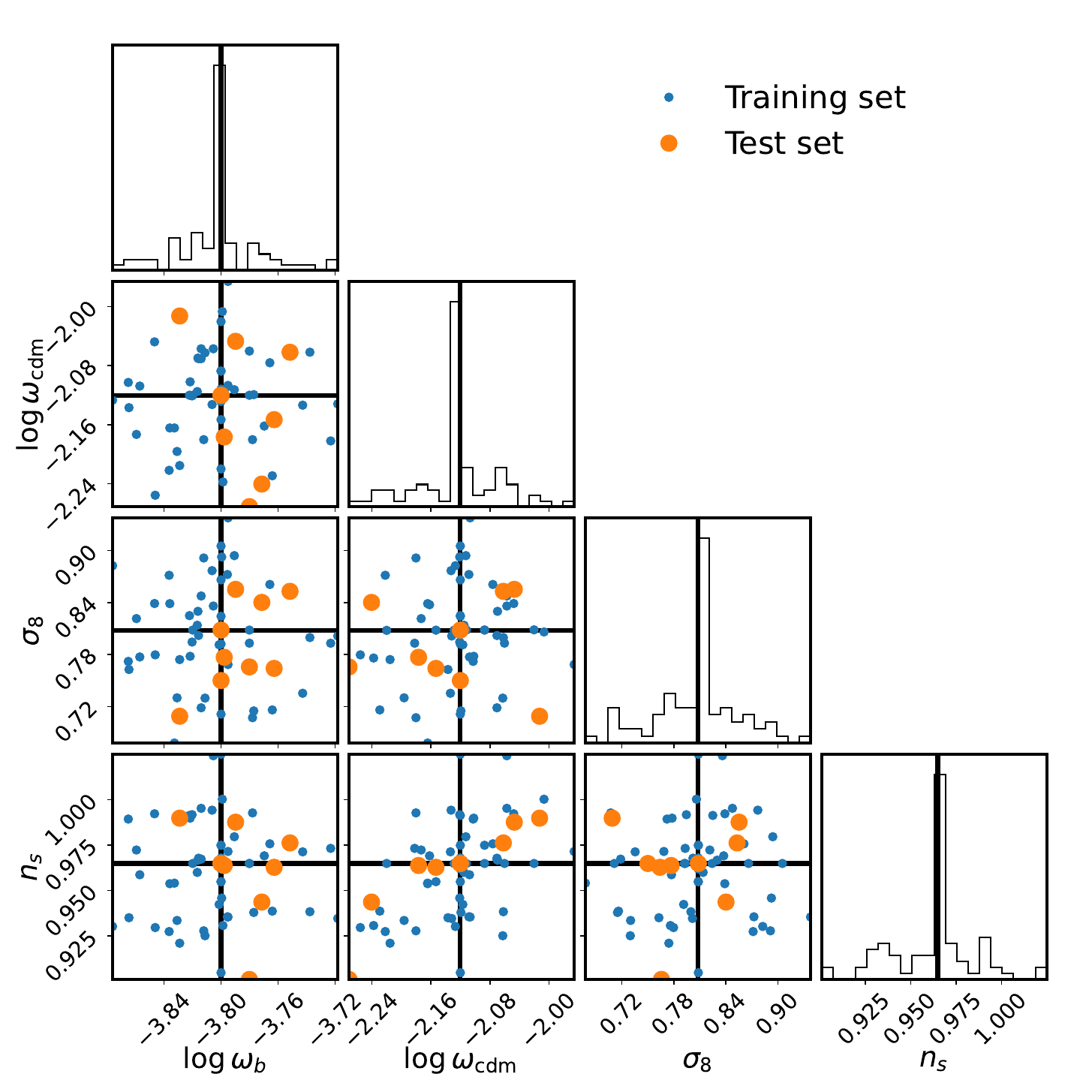}
    \caption{Training and test cosmologies in $\Lambda$CDM parameter space. Top panels in each row show the distribution of each parameter. Note that we omit $h$ as it is fixed to the acoustic scale.}
    \label{fig:traintest}
\end{figure}

Here we mainly use the \textsc{AbacusSummit} cosmology grid, a set of 85 base boxes run at 85 distinct cosmologies, at fixed initial phase. The 85 cosmologies are tagged c000-181 non-consecutively and visualised in Figure~\ref{fig:traintest} (see \citealt{2021Maksimova} for details). The details of each cosmology are described on the \textsc{AbacusSummit} website. \footnote{\url{https://abacussummit.readthedocs.io/en/latest/cosmologies.html}}

While we only conduct our analysis in $\Lambda$CDM cosmology space in this paper, we train the emulator in a larger $w$CDM+$N_\mathrm{eff}$+running space with 8 parameters: the baryon density $\omega_b = \Omega_b h^2$, the cold dark matter density $\omega_\mathrm{cdm} = \Omega_\mathrm{cdm}h^2$, the amplitude of structure $\sigma_8$, the spectral tilt $n_s$, running of the spectral tilt $\alpha_s$, the density of massless relics $N_\mathrm{eff}$, and dark energy equation of state parameters $w_0$ and $w_a$ ($w(a) = w_0+(1-a)w_a$). 

A set of observational constraints were followed in the design of the parameter grid. For example, $w_a$ is varied while holding the equation of state
at $z = 0.333$ constant, so that the low-redshift cosmic distance scale is minimally changed. Similarly, changes in $N_\mathrm{eff}$ are compensated by changes in $\omega_\mathrm{cdm}$ and $n_s$, to minimise changes in the CMB. A flat spatial curvature is assumed for all cosmologies, and the Hubble constant $H_0$ is chosen to match the CMB acoustic scale $\theta_*$ as measured by Planck 2018 (P18) cosmology \citep{2020Planck}. We briefly summarise the cosmology choices here. 

\textsc{c000} corresponds to the fiducial P18 cosmology, specifically the mean estimates of the Planck TT,TE,EE+lowE+lensing likelihood chains. The 25 boxes with different phases are at \textsc{c000}.

\textsc{c001-004} correspond to four secondary cosmologies. Specifically, a WMAP9 + ACT + SPT \citep[][]{2017Calabrese}, a thawing dark energy model ($w_0 = -0.7, w_a = -0.5$), a model with extra relativistic density ($N_\mathrm{eff} = 3.7$), and a model with lower amplitude of clustering (c000 but with $\sigma_8$ = 0.75). 

\textsc{c100-126} is a linear derivative grid set up around \textsc{c000}, with symmetric pairs along all 8 parameter axes. The grid also includes additional pairs along $\omega_\mathrm{cdm}$, $\sigma_8$, $n_s$, $w_0$, and $w_a$ with smaller step sizes. 

\textsc{c130-181} forms an emulator grid that provides a wider coverage of the 8-dimensional parameter space. This was done by placing electrostatic points on the surface of an 8-dimensional ellipsoid, whose extent was chosen to be 3 to 8 standard deviations beyond current constraints from the combination of CMB and large-scale structure data. The grid excludes anti-podal reflected points and includes extra excursion along $\sigma_8$.

These 85 cosmologies span the $w$CDM+$N_\mathrm{eff}$+running parameter space and form the basis of our simulation-based model. It is worth pointing out that while we train our emulators over this broader cosmology model space, the analysis presented in the rest of this paper only considers the $\Lambda$CDM space, fixing the other parameters to their Planck2018 values. Additionally, the neutrino mass is fixed at 60meV and the Hubble parameter $h$ is fully degenerate with matter density by fixing the acoustic scale across all 85 cosmologies. We refer the readers to \citealt{2021Maksimova} for full description and justification of the cosmology design. 

\subsection{The Halo Occupation Distribution (HOD)}
\label{subsec:model}
The galaxy--halo connection model we use for generating the realistic mocks and for the forward model is known as the Halo Occupation Distribution \citep[HOD; e.g.][]{2005Zheng, 2007bZheng}, which probabilistically populates dark matter haloes with galaxies according to a set of halo properties. For a Luminous Red Galaxy (LRG) sample, the HOD is well approximated by a vanilla model given by:
\begin{align}
    \bar{n}_{\mathrm{cent}}^{\mathrm{LRG}}(M) & = \frac{f_\mathrm{ic}}{2}\mathrm{erfc} \left[\frac{\log_{10}(M_{\mathrm{cut}}/M)}{\sqrt{2}\sigma}\right], \label{equ:zheng_hod_cent}\\
    \bar{n}_{\mathrm{sat}}^{\mathrm{LRG}}(M) & = \left[\frac{M-\kappa M_{\mathrm{cut}}}{M_1}\right]^{\alpha}\bar{n}_{\mathrm{cent}}^{\mathrm{LRG}}(M),
    \label{equ:zheng_hod_sat}
\end{align}
where the five vanilla parameters characterizing the model are $M_{\mathrm{cut}}, M_1, \sigma, \alpha, \kappa$. The parameter \sandy{$M_{\mathrm{cut}}$ sets the halo mass at which the mean central occupation is 0.5}, while $M_1$ characterises the typical halo mass that hosts one satellite galaxy. $\sigma$ describes the steepness of the transition from 0 to 1 in the number of central galaxies, $\alpha$ is the power law index on the number of satellite galaxies, and $\kappa M_\mathrm{cut}$ gives the minimum halo mass to host a satellite galaxy.
We have added a modulation term $\bar{n}_{\mathrm{cent}}^{\mathrm{LRG}}(M)$ to the satellite occupation function to remove satellites from haloes without centrals\footnote{There is evidence that such central-less satellites may exist in a realistic stellar-mass selected catalogue \citep[][]{2019Jimenez}. We include this term for consistency with previous works, but it should have minimal impact on clustering.}. We have also included an incompleteness parameter $f_\mathrm{ic}$, which is a downsampling factor controlling the overall number density of the mock galaxies. This parameter is relevant when trying to match the observed mean density of the galaxies in addition to clustering measurements. By definition, $0 < f_\mathrm{ic}\leq 1$.

In addition to determining the number of galaxies per halo, the standard HOD model also dictates the position of velocity of the galaxies. For the central galaxy, its position and velocity are set to be the same as those of the halo center, specifically the L2 subhalo center-of-mass for the {\sc CompaSO} haloes. For the satellite galaxies, they are randomly assigned to halo particles with uniform weights, each satellite inheriting the position and velocity of its host particle. 

To model redshift-space distortion, we also include an additional HOD extension known as velocity bias, which biases the velocities of the central and satellite galaxies. This is shown to to be a necessary ingredient in modeling BOSS LRG redshift-space clustering on small scales \citep[e.g.][]{2015aGuo, 2021bYuan}. Velocity bias has also been identified in hydrodynamical simulations and measured to be consistent with observational constraints \citep[e.g.][]{2017Ye, 2022Yuan}. 

\sandy{We parametrise velocity bias through two additional parameters: \texttt{$\alpha_\mathrm{vel, c}$} is the central velocity bias parameter, which modulates the peculiar velocity of the central galaxy relative to the halo center along the line-of-sight (LoS). Specifically in this model, the central galaxy velocity along the LoS is thus given by 
    \begin{equation}
        v_\mathrm{cent, z} = v_\mathrm{L2, z} + \alpha_\mathrm{vel, c} \delta v(\sigma_{\mathrm{LoS}}),
        \label{equ:alphac}
    \end{equation}
    where $v_\mathrm{L2, z}$ denotes the LoS component of the central subhalo velocity, $\delta v(\sigma_{\mathrm{LoS}})$ denotes the Gaussian scatter, and $\alpha_\mathrm{vel, c}$ is the central velocity bias parameter. By definition, $\alpha_\mathrm{vel, c} = 0$ corresponds to no central velocity bias. We also define $\alpha_\mathrm{vel, c}$ as non-negative, as negative and positive $\alpha_c$ are fully degenerate observationally. 
}

\sandy{The second parameter is \texttt{$\alpha_\mathrm{vel, s}$}, the satellite velocity bias parameter, which modulates how the satellite galaxy peculiar velocity deviates from that of the local dark matter particle. 
    Specifically, the satellite velocity is given by 
    \begin{equation}
        v_\mathrm{sat, z} = v_\mathrm{L2, z} + \alpha_\mathrm{vel, s} (v_\mathrm{p, z} - v_\mathrm{L2, z}),
        \label{equ:alpha_s}
    \end{equation}
    where $v_\mathrm{p, z}$ denotes the line-of-sight component of particle velocity, and $\alpha_\mathrm{vel, s}$ is the satellite velocity bias parameter.
    $\alpha_\mathrm{vel, s} = 1$ indicates no satellite velocity bias, i.e. satellites perfectly track the velocity of their underlying particles. 
}

So far, the baseline HOD is parameterised by 8 parameters, $M_{\mathrm{cut}}, M_1, \sigma, \alpha, \kappa$, $\alpha_\mathrm{vel, c}$, $\alpha_\mathrm{vel, s}$, and $f_\mathrm{ic}$. We now introduce three physically motivated extensions in addition to the baseline HOD:
\begin{itemize}

    \item \texttt{$A_\mathrm{cent}$} or \texttt{$A_\mathrm{sat}$} are the concentration-based secondary bias parameters for centrals and satellites, respectively. Also known as galaxy assembly bias parameters. $A_\mathrm{cent} = 0$ and $A_\mathrm{sat} = 0$ indicate no concentration-based secondary bias in the centrals and satellites occupation, respectively. A positive $A$ indicates a preference for lower-concentration haloes, and vice versa. 
    \item \texttt{$B_\mathrm{cent}$} or \texttt{$B_\mathrm{sat}$} are the environment-based secondary bias parameters for centrals and satellites, respectively. The environment is defined as the mass density within a $r_\mathrm{env} = 5h^{-1}$Mpc tophat of the halo center, excluding the halo itself. $B_\mathrm{cent} = 0$ and $ B_\mathrm{sat} = 0$ indicate no environment-based secondary bias. A positive $B$ indicates a preference for haloes in less dense environments, and vice versa. 
    \item \texttt{$s$} is the baryon feedback parameter, which modulates how the radial distribution of satellite galaxies within haloes deviates from the radial profile of the halo (mimicking baryonic effects). $s = 0$ indicates no radial bias, i.e. satellites follow the dark matter distribution. $s > 0$ indicates a more extended (less concentrated) profile of satellites relative to the dark matter, and vice versa. It has also been shown that galaxy selection based on luminosity and SFR can bias the satellite profile \citep{2018Orsi}. Thus, the $s$ parameter can additionally marginalise over some selection effects.
\end{itemize}

Throughout the rest of this paper, we conduct fits with three different HOD models, each including a subset of the three extensions. To keep future participants blind, we refer to these models as models A, B, and C. A key objective of this paper is to compare these three different models and demonstrate that one model is validated by the data while the other two are not (section~\ref{subsec:hodval}). \sandy{For completeness, we also tested a fourth model that includes all aforementioned extensions, but we do not see any improvements in goodness-of-fits and model evidence so we omit that from this paper. It is important to keep the HOD model compact so as to not dilute the cosmological information in the data. }

For computational efficiency, we adopt the highly optimised \ahod\ implementation, which significantly speeds up the HOD calculation per HOD parameter combination \citep[][]{2021bYuan}. The code is publicly available as a part of the \textsc{abacusutils} package at \url{https://github.com/abacusorg/abacusutils}. Example usage can be found at \url{https://abacusutils.readthedocs.io/en/latest/hod.html}.

\subsection{2D-$k$NN}
\label{sec:statistics}

\sandy{The $k$-th nearest neighbor ($k$NN) are highly informative statistics that summarise the spatial distribution of data points by tabulating their distances to a set of random or uniform volume-filling query points. \cite{2021Banerjee, 2021bBanerjee} showed through Fisher forecasts that $k$NNs are highly informative on cosmological parameters and are sensitive to all orders of $n$-point correlation functions. Operationally, for each $k$, we identify each query point's $k$-th nearest data points and record their distances $r_i^k$, where $i$ loops through all the query points and $k$ represents the order. These $r_i^k$s can then be summarised with a cumulative distribution function (CDF) as a function of length. The ensemble of these CDFs for all $k$s forms our $k$NN summary statistics.}

\begin{figure*}
    \centering
    \hspace*{-0.5cm}
    \includegraphics[width = 6in]{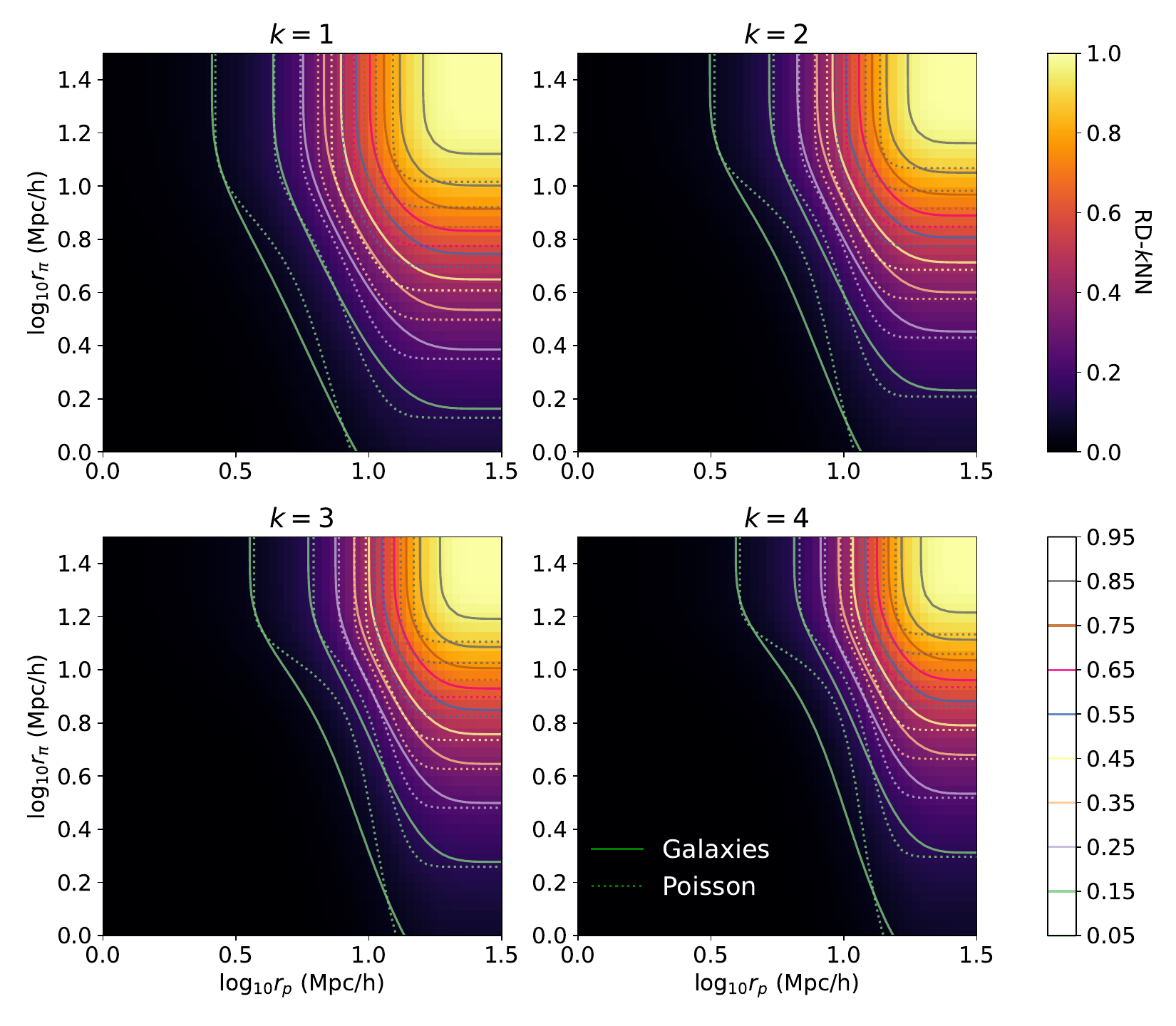}
    \caption{RD-$k$NN statistics calculated on the redshift-space Beyond-2p mock, averaged over 10 realizations. Each panel corresponds to a $k$ order. In each panel, the colour gradient represents the CDF as a function of projected separation and line-of-sight separation. The solid lines show the contours of the CDF. The dotted lines show the contours of a CDF of an un-clustered Poisson random sample. The difference between the solid and dotted contours indicates the signatures of clustering. We show only the first 4 $k$ orders for brevity.}
    \label{fig:knns_target_RD}
\end{figure*}
\begin{figure*}
    \centering
    \hspace*{-0.5cm}
    \includegraphics[width = 6in]{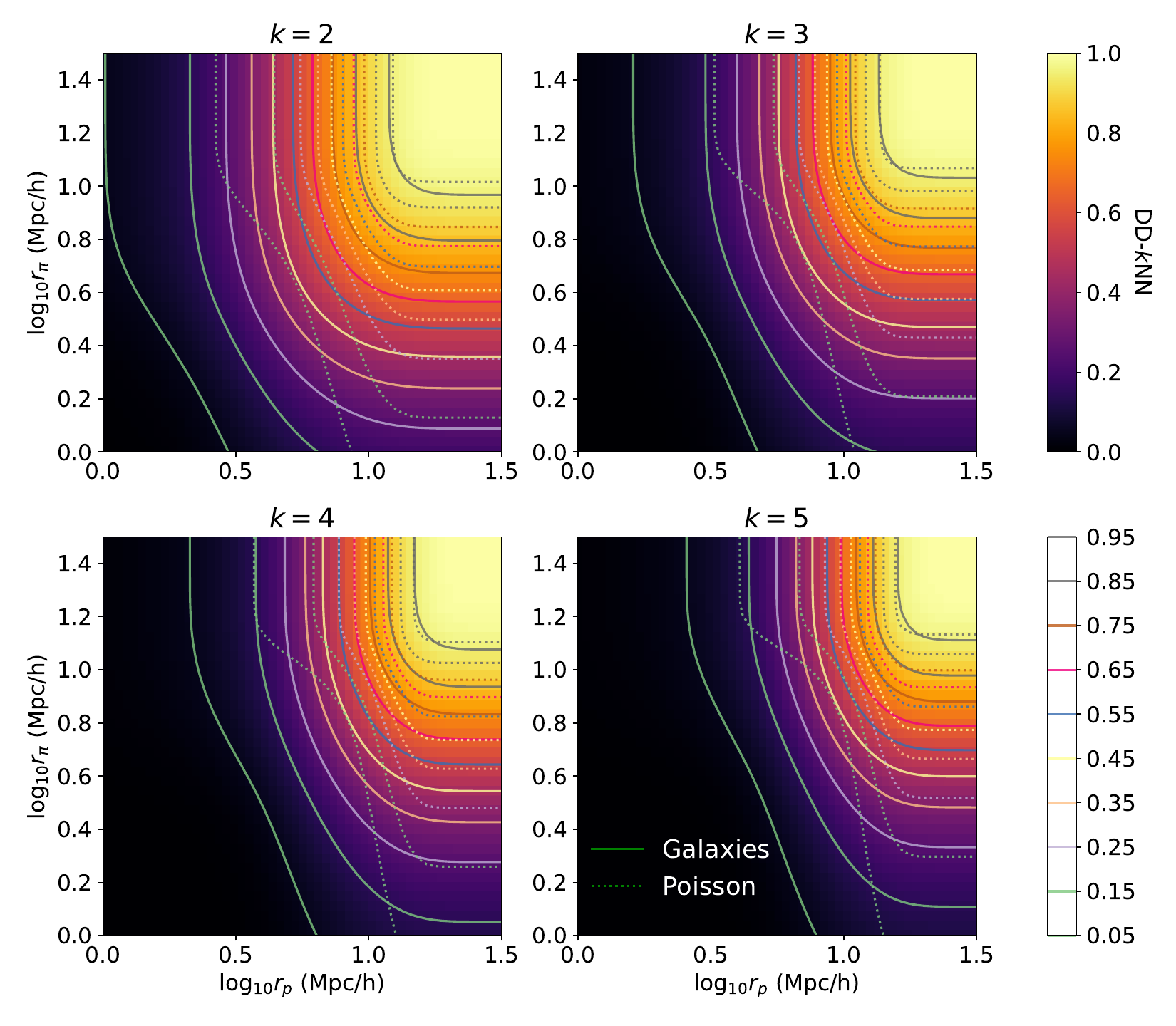}
    \caption{Visualizations of the DD-$k$NN statistics calculated on the redshift-space I mock, averaged over 10 realizations. The lines are colours are definely similarly to Figure~\ref{fig:knns_target_RD}. Note that we have defined DD-$1$NN~$=1$, thus $k = 2$ is the first meaningful order. }
    \label{fig:knns_target_DD}
\end{figure*}

\cite{2023Yuan} generalised the standard $k$NN to 2D to disentangle the redshift-space distortion features from the projected galaxy clustering. 
Specifically, we decompose the distance between each query--data pair $r$ into a $r_p$ and a $r_\pi$ component, and we bin both projections into a 2D histogram. Then we calculate a 2D CDF where each bin accumulates the counts from all bins with smaller $r_p$ and $r_\pi$. Finally, we normalise the cumulative counts by the total number of query points. Conceptually, the 2D $k$NN-CDF is exactly analogous to the default 1D $k$NN-CDF, except we tabulate distances and the cumulative statistics in 2D. \cite{2023Yuan} showed that the 2D $k$NNs are richly informative and fully derive other summary statistics such as the 2-point correlation function (2PCF), counts-in-cells, and the void probability function (VPF). The study also showed that the 2D $k$NNs place tight constraints on HOD parameters in a realistic setting. 

For this analysis, we adopt the 2D $k$NN statistics to comprehensively extract information about galaxy density and clustering on non-linear scales. Specifically, we separately analyze both the query--data $k$NN (RD-$k$NN) and the data--data $k$NN (DD-$k$NN). \sandy{\cite{2023Yuan} described two flavors of $k$NNs in detail, but we briefly describe them here. RD-$k$NN represents the standard setup, where we put down a volume-filling set of uniform or random query points and we tabulate their distances to the data points. With DD-$k$NN, we use the data points as query points, thus we tabulate the distance between data points and data points. }

Although the two statistics might seem similar at first glance, they reveal distinct facets of the galaxy field. The RD-$k$NN is primarily a density statistic, capturing density counts in specific volumes (counts-in-cells) and detailing the size and distribution of voids. In contrast, the DD-$k$NN enumerates the galaxy pairs in group and cluster regions. It also serves as a generating function for the 2PCF and closely relates to high-order correlation functions. The unique information provided by the two statistics becomes apparent in their respective constraints. We also conduct cross-validation tests between the two statistics to assess the robustness of our results. For computational efficiency, we adopt the parallel $k$NN implementation written in \textsc{Rust}\footnote{\url{https://crates.io/crates/fnntw/}} and wrapped for \textsc{Python}\footnote{\url{https://pypi.org/project/pyfnntw/}}. 

We set up our 2D $k$NN data vector as follows: we use 8 logarithmic bins along the $r_p$ direction between $0.32h^{-1}$Mpc and $63h^{-1}$Mpc, and 5 logarithmic bins along the $r_\pi$ direction between $1h^{-1}$Mpc and $32h^{-1}$Mpc. We include $k = 1,2,3,..., 9$. For RD-$k$NN, we set up the query points as a uniformly spaced grid with cell length $10h^{-1}$Mpc. We also remove bins where the CDF values are less than 0.05 or greater than 0.95 as those bins are noisy and contain little physical information. As a result, the DD-$k$NN data vector is of length 143 whereas the RD-$k$NN data vector is of length 114. The minimum transverse scale after removing these bins is approximately 3$h^{-1}$Mpc for RD-$k$NN and 0.5$h^{-1}$Mpc for DD-$k$NN. 

We justify our scale choices as follows. Our methodology is focused on non-linear scales. The maximum scale in projected separation is chosen to obtain some power from linear scales but not rely on well-known large-scale features such as the BAO. The modest maximum projected separation also reduces any potential biases in the jackknife covariance calculation. The maximum separation along the line of sight is designed to capture the finger-of-god and large-scale Kaiser effects. 
The fiducial minimum scale reaches into the 1-halo regime, which is well measured given the volume but which we expect to be significantly affected by halo definition and baryon feedback. After removing noisy bins, the RD-$k$NN should be largely insensitive to 1-halo scale physics, but the DD-$k$NN remains sensitive to very small scales. Thus, it is essential that we validate our modeling of the smallest scales for this statistic. 

Figure~\ref{fig:knns_target_RD} and Figure~\ref{fig:knns_target_DD} visualise the 2D $k$NN statistics up to the first four orders. We show the measurement on the redshift-space mocks as well as the measurement on an unclustered Poisson random sample of the same size for comparison. The difference between the solid and dashed contours denotes the informative features in the statistics. 
The corresponding covariance matrices are calculated from 1250 jackknife volumes cut from the provided boxes. Each box is of length $400h^{-1}$Mpc, sufficient for the scales considered. The covariance matrices are visualised in Figure~\ref{fig:knns_cov_rd} and Figure~\ref{fig:knns_cov_dd}, respectively. 

\begin{figure}
    \centering
    \hspace*{-0.5cm}
    \includegraphics[width = 3.3in]{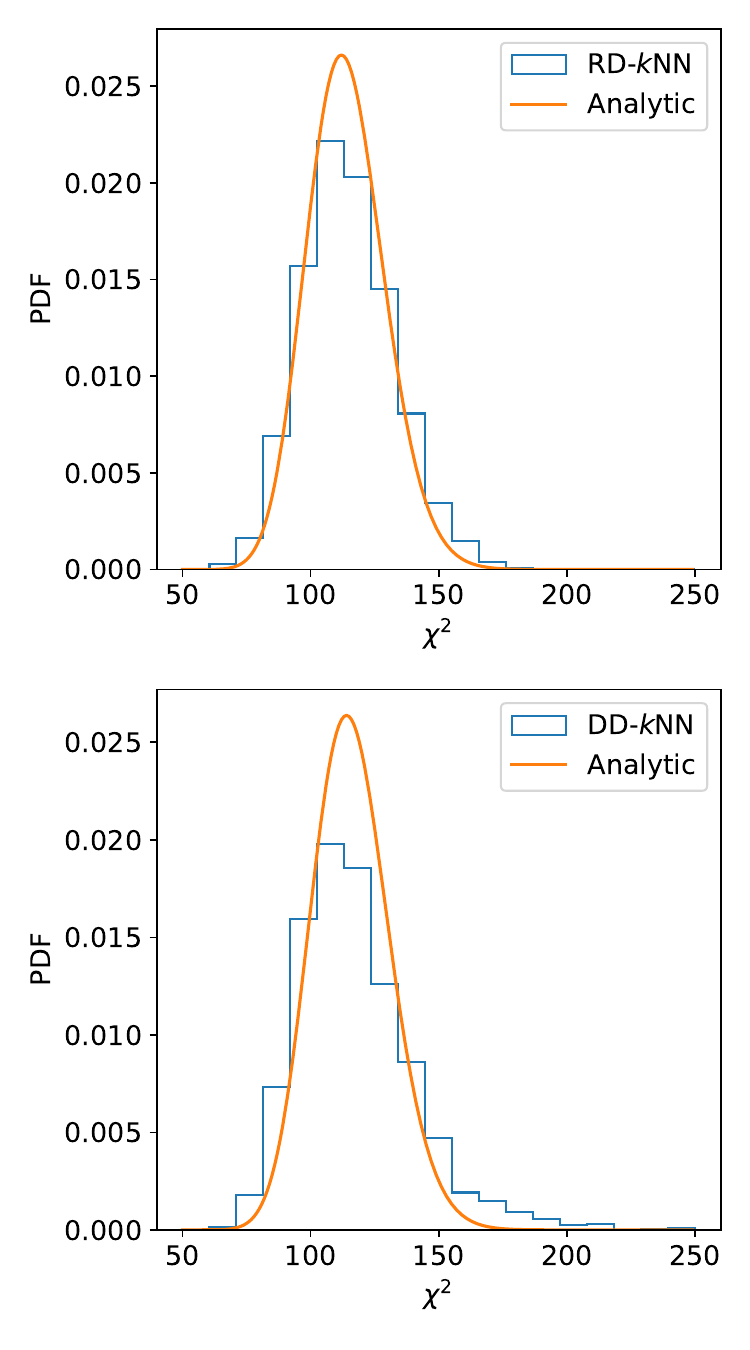}
    \vspace{-0.5cm}
    \caption{A qualitative assessment of the Gaussianity of the likelihoods for RD-$k$NN (top panel) and DD-$k$NN (bottom panel). The histograms show the distribution of $\chi^2$ values as measured from 1883 \textsc{AbacusSummit} small boxes. The solid line shows an analytic $\chi^2$ distribution with degrees of freedom set to the number of bins, as we expect from Gaussian statistics. Both statistics are consistent with the analytic distributions and with our assumption of a Gaussian likelihood. }
    \label{fig:chi2test}
\end{figure}

For comparison, we also make several references to the 2-point correlation function (2PCF) in this analysis. Specifically, we make use of the redshift-space 2PCF $\xi(r_p, r_\pi)$, which can be computed using the \citet{1993Landy} estimator:
\begin{equation}
    \xi(r_p, r_\pi) = \frac{DD - 2DR + RR}{RR},
    \label{equ:xi_def}
\end{equation}
where $DD$, $DR$, and $RR$ are the normalised numbers of data--data, data--random, and random--random pair counts in each bin of $(r_p, r_\pi)$; $r_p$ and $r_\pi$ are the transverse and line-of-sight separations in comoving units. The redshift space 2PCF $\xi(r_p, r_\pi)$ in principle represents the full information content of the 2PCF.

For the rest of the paper, we adopt a Gaussian likelihood function for both $k$NN statistics, specifically 
\begin{equation}
    \log L =  -\frac{1}{2}(x_\mathrm{proposed} - x_\mathrm{target})^T \boldsymbol{C}^{-1}(x_\mathrm{proposed} - x_\mathrm{target}),
\end{equation}
where $x$ is the $k$NN data vector and $\boldsymbol{C}$ is the covariance matrix. We do not include a mean density term in the likelihood as the mean density information is naturally captured by the $k$NNs.

We test the assumption of a Gaussian likelihood as follows. If the likelihood of the summary statistic is Gaussian distributed, the $\chi^2$ values should also follow a $\chi^2$ distribution with degrees of freedom determined by the number of bins.
Figure~\ref{fig:chi2test} shows the comparison of 1883 independent realizations of mock $k$NN statistics calculated on the \textsc{AbacusSummit} small boxes with analytic $\chi^2$ distributions. Both our statistics are consistent with a  $\chi^2$ distribution and thus our Gaussian assumptions are valid. There is a high-$\chi^2$ tail that exceeds the analytic prediction in the DD-$k$NN case. Additional scale cuts, employed in the following sections, remove this excess. 

\subsection{Forward model and emulator}
\label{subsec:emulator}

\sandy{Having defined the summary statistics and likelihood function, we now describe our modeling methodology that predicts the relevant summary statistics given arbitrary input parameters in cosmology and HOD. 
Specifically, we use a forward model as follows.} Starting from dark matter only simulations parameterised with cosmological parameters, we populate the simulated haloes with galaxies using parameterised HOD models. Then we compute the $k$NN statistics on the resulting mock galaxy density. We compute the likelihood function by comparing the mock predicted $k$NN with the measured $k$NN on the target sample, incorporating the jackknife covariance matrix and assuming a Gaussian likelihood function. 

Due to the high computational cost of running large, high-resolution simulations, we cannot run arbitrary cosmologies on the fly, but instead, we have to rely on the 85 cosmologies in the \textsc{AbacusSummit} suite to build an approximate emulator model. To achieve this, we populate each cosmology box with a few thousand of HOD models and record the resulting $k$NNs to form a large training set. Specifically, we follow the approach of \cite{2022bYuan}, where we take advantage of the high efficiency of the \ahod\ code and run MCMC chains in the HOD parameter space against the target data vector at each cosmology. We stop the MCMC chains after 20,000 evaluations in each box and select samples whose likelihood is greater than $\log L > -9000$. This approach limits the subsequent emulator training to a compact region in the cosmology+HOD parameter space, reducing the effects of outliers and improving the emulator precision. 

Having selected the training sample, we now build an emulator or surrogate model that takes the cosmology and HOD parameters as inputs and outputs the desired $k$NN statistics as outputs. For the emulator model, we adopt a fully connected neural network of 5 layers and 500 nodes per layer with Randomised Leaky Rectified Linear Units (RReLU) activation. We train the network following a mini-batch routine with the Adam optimiser and a mean squared loss function, where we use the diagonal terms of the covariance matrix as bin weights.

To test the performance of the emulator, we remove 9 cosmologies (\textsc{c001-004} and \textsc{c171-175}) from the training set and reserve them as outsample tests. Figure~\ref{fig:traintest} visualises the distribution of the training and test sets in $\Lambda$CDM parameter space. At each of the 9 cosmologies, we randomly sample 400 HODs with $\log L > -9000$ and combine them to form a test set of size 3600. We compute the mean absolute error of the emulator on all $9\times 400$ test samples and find a mean error that is sub-dominant to the data error, approximately 30--60$\%$ of the jackknife error. We further conduct tests to ensure the emulator is not suffering from systematic biases towards the mean (see Figure~10 of \citealt{2022cYuan}). Finally, we summarise the emulator errors by computing an emulator covariance matrix from the hold-out test samples and add it to our jackknife covariance matrices. We provide the emulator correlation matrices in Figure~\ref{fig:corr_dd_emu} and Figure~\ref{fig:corr_rd_emu}. The final covariance matrices going into our likelihood function is a combination of data sample variance ($\mathbf{C^\mathrm{jackknife}}$, accounting for the phase difference between the data and the model) and emulator errors ($\mathbf{C^\mathrm{emulator}}$):
\begin{equation}
    \mathbf{C^\mathrm{comb}} = \mathbf{C^\mathrm{jackknife}} + \mathbf{C^\mathrm{emulator}}
\end{equation}

To summarise, our model starts from the \textsc{AbacusSummit} simulations, and we model galaxy distribution with \textsc{AbacusHOD}, invoking several extensions to the standard HOD model. We then compute the desired summary statistics and emulate that as a function of cosmology and HOD using a neural net model. We finally confront our forward model with the mock data by sampling the parameter posteriors and deriving constraints. 

\section{Results}
\label{sec:results}
In this section, we present the ensemble results with different analysis choices and then describe the validation tests and model selection tests carried out to identify our final results presented in \cite{b2p}. We adopt flat priors for each parameter, and Table~\ref{tab:knn_priors} summarises the prior bounds. For posterior sampling, we employ the efficient nested sampling package \textsc{dynesty} \citep{2018Speagle, 2019Speagle}. We initiate each nested sampling chain with 2000 live points and a stopping criterion of $d\log\mathcal{Z} = 0.01$, where $\mathcal{Z}$ is the evidence.

\begin{table}
    \centering
    \begin{tabular}{lcc}
        \hline
        \hline
        Parameter & Bounds         \\
        \hline
    $\ln \omega_b$  & [-4.5, -2] \\
    $\ln \omega_\mathrm{cdm}$ & [-3, -1] \\  
    $\sigma_8$   &  [0.5, 1.1]  \\
    $ n_s$  &    [0.8, 1.2] \\
        \hline
        $\log_{10} M_\mathrm{cut}$   & $[12.0, 14.5]$  \\
        $\log_{10} M_1$              & $[13.0, 16.0]$\\
        $\log_{10}\sigma$            & $[-3.5, 1.5]$  \\
        $\alpha$                & $[0.5, 1.5]$ \\
        $\kappa$                & $[0.0,2.0]$ \\
        $\alpha_c$     & [0.0, 1.0] \\
        $\alpha_s$     & [ 0.2, 1.8] \\
        \hline
        $B_\mathrm{cent}$       & $[-1.0,1.0]$ \\
        $B_\mathrm{sat}$       & $[-1.0,1.0]$  \\
        $s$       & $[-1.0,1.0]$  \\
        \hline        
    \end{tabular}
    \caption{Prior bounds for the $k$NN analyses, for the cosmology parameters, HOD parameters, and assembly bias parameters.}
    \label{tab:knn_priors}
\end{table}

We note that all analyses in this section were conducted blind and all the results were finalised before unblinding. We do not present or make reference to any analysis done post-unblinding. We also do not omit blind analyses that might be deemed unsatisfactory post unblinding. However, the texts presented in this section were written post-unblinding and make reference to the true values for the sake of discussion. 

\subsection{Cosmology posteriors}
\label{subsec:posteriors}
\begin{figure*}
    \hspace{-0.3cm}
    \includegraphics[width=0.7\textwidth]{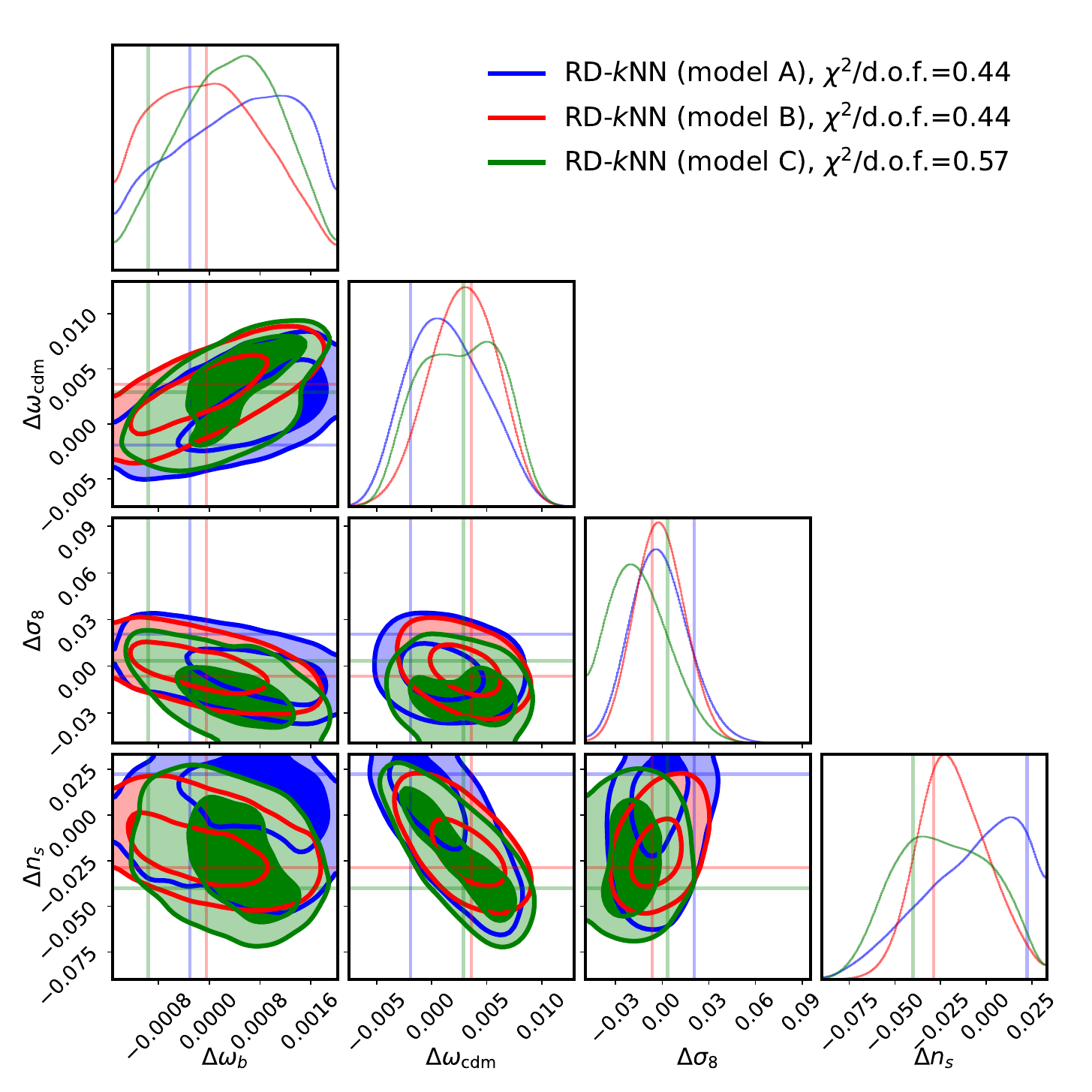}
    \vspace{-0.3cm}
    \caption{The $\Lambda$CDM cosmology posteriors inferred with the RD-$k$NN data vector using three different HOD models. The axes show the difference between the inferred and true values. The three HOD models are summarised in section~\ref{subsec:model}. For the RD-$k$NN data vector, we use $k = 1,2,3,...,9$, covering scales $3h^{-1}\mathrm{Mpc} < r_p < 63h^{-1}\mathrm{Mpc}$ and $0.5h^{-1}\mathrm{Mpc} < r_\pi < 31.6h^{-1}\mathrm{Mpc}$. The fainter solid lines correspond to maximum likelihood points. We omit $h$ as it is fixed to the acoustic scale. The three contours show good consistency with each other and the input values. \sandy{The contours correspond to $68\%$ and $95\%$ confidence intervals.}}
    \label{fig:corner_rdknn}
\end{figure*}

Figure~\ref{fig:corner_rdknn} presents the 2D marginalised posteriors when fitting the RD-$k$NN in the $\Lambda$CDM cosmology space. Note that for this analysis, we fix $w_0 = -1$ and $w_a = 0$. The Hubble parameter $h$ is fully degenerate with matter density by fixing the acoustic scale.  The three colours correspond to constraints from the different HOD models. The legend summarises the goodness-of-fits, and the contours show the 1-2$\sigma$ constraints. 

We see that all three models correctly recover the true cosmology and derive mutually consistent cosmology. This shows that the RD-$k$NN is a powerful statistic that captures non-linear information without being necessarily sensitive to the details of HOD modeling. This is perhaps not surprising as we have configured it with a query spacing of $10h^{-1}$Mpc, thus making it mostly sensitive to densities on scales of a few Megaparsecs or larger and removing much of the sensitivity to 1-halo physics. It is also worth reminding that the RD-$k$NN is a density statistic that measures the configuration of mass around volume-sampled queries, meaning that it is more sensitive to the mass distribution around voids. This distinguishes RD-$k$NN from clustering statistics (DD-$k$NN and 2PCF) that measure galaxy--galaxy distances, which sample mostly highly clustered regions that are strongly affected by 1-halo physics. 

\begin{figure*}
    \hspace{-0.3cm}
    \includegraphics[width=0.7\textwidth]{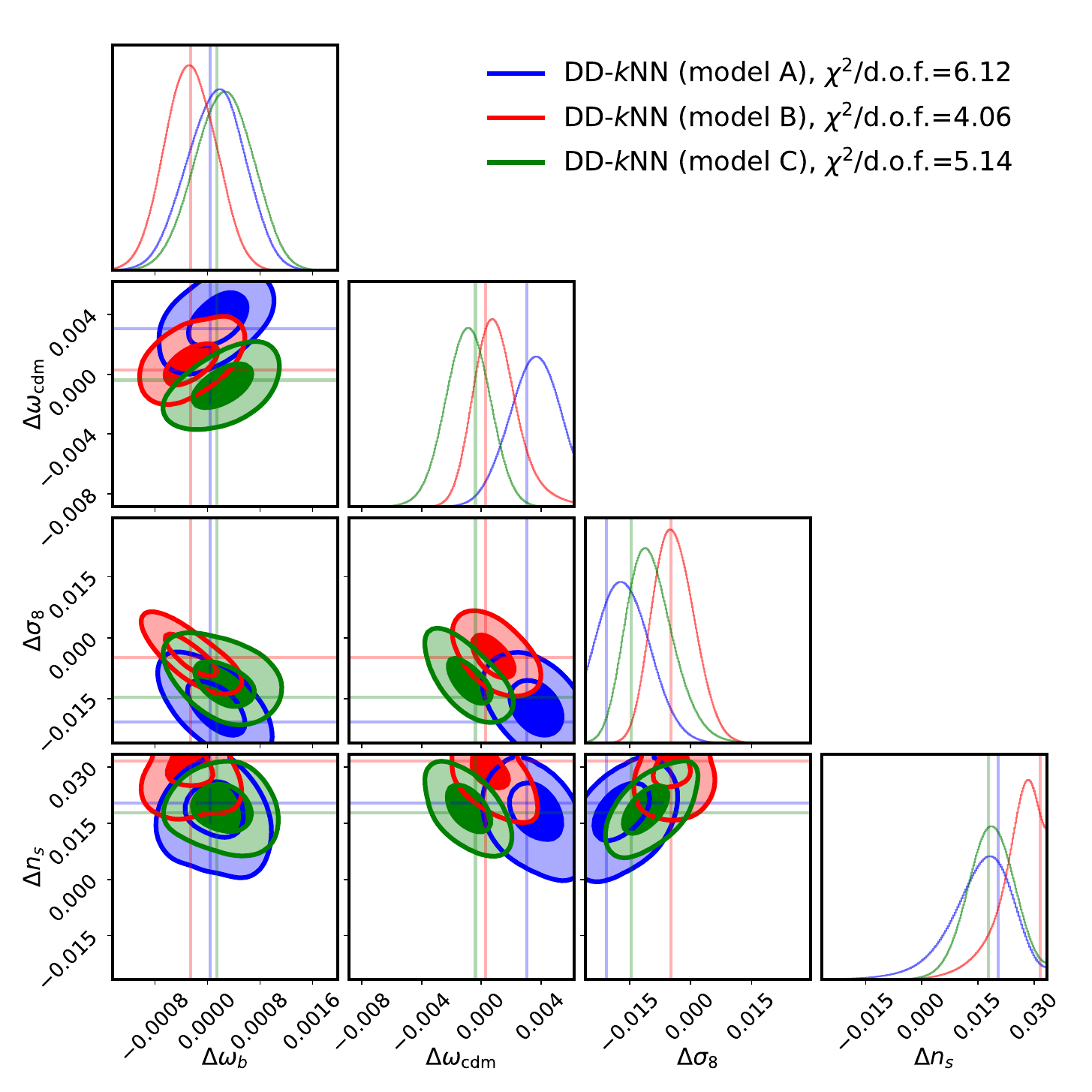}
    \vspace{-0.3cm}
    \caption{The $\Lambda$CDM cosmology posteriors inferred with the DD-$k$NN data vector using three different HOD models. The axes show the difference between the inferred and true values. For the DD-$k$NN data vector, we use $k = 2,3,...,9$, covering scales $0.67h^{-1}\mathrm{Mpc} < r_p < 63h^{-1}\mathrm{Mpc}$ and $0.5h^{-1}\mathrm{Mpc} < r_\pi < 31.6h^{-1}\mathrm{Mpc}$. The three contours show mild disagreement at the 2-3$\sigma$ level. The goodness-of-fit as shown in terms of $\chi^2$/d.o.f. is poor. \sandy{The contours correspond to $68\%$ and $95\%$ confidence intervals.} }
    \label{fig:corner_ddknn}
\end{figure*}

In Figure~\ref{fig:corner_ddknn}, we show the cosmology posteriors obtained from fitting the DD-$k$NN. 
The three models are mutually in up to $2\sigma$ tension with each other. The large goodness-of-fit values indicate that our model is not flexible enough to produce the features in the DD-$k$NN. Given these discrepancies, we either need more flexible models or must apply further scale cuts to achieve reliable results. Note that this decision was made before unblinding, and was based primarily on observed tensions among the models and the poor $\chi^2$ value. Without access to additional information about the target galaxy sample --- such as spectral energy distributions, selection criteria, or imaging data --- that would typically be available in a real survey, we do not have straightforward ways to decouple the galaxy formation model from cosmology. Hence, we choose to use additional scale cuts to improve the fit.

\begin{figure}
    \hspace{-0.3cm}
    \includegraphics[width=0.45\textwidth]{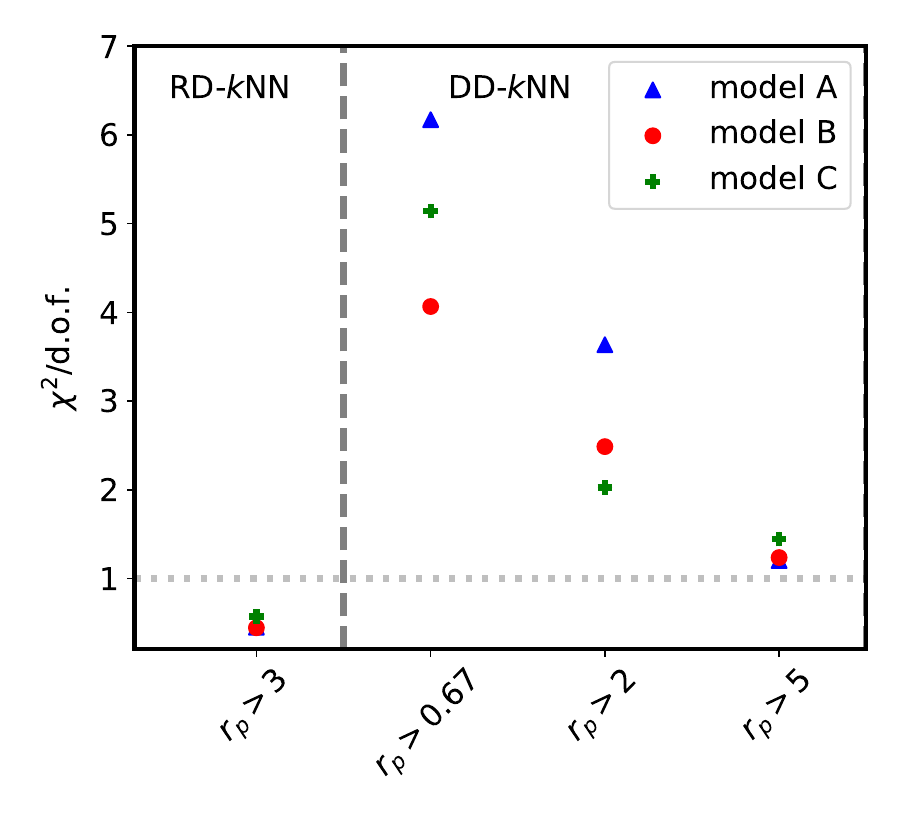}
    \vspace{-0.3cm}
    \caption{The best-fit $\chi^2$/d.o.f. when using different minimum scale cuts. Different colours refer to the three different HOD models. The $x$-axis iterates through different scale cuts (in units of $h^{-1}$Mpc), showing the fits grouped vertically by data vector. DD-$k$NN only achieves a good fit at $r_p > 5h^{-1}$Mpc; RD-$k$NN achieves a good fit at smaller scales. This highlights the DD-$k$NN's high sensitivity to galaxy bias and the details of the HOD model on small scales. }
    \label{fig:chi2_scale}
\end{figure}

\begin{figure*}
    \hspace{-0.3cm}
    \includegraphics[width=0.7\textwidth]{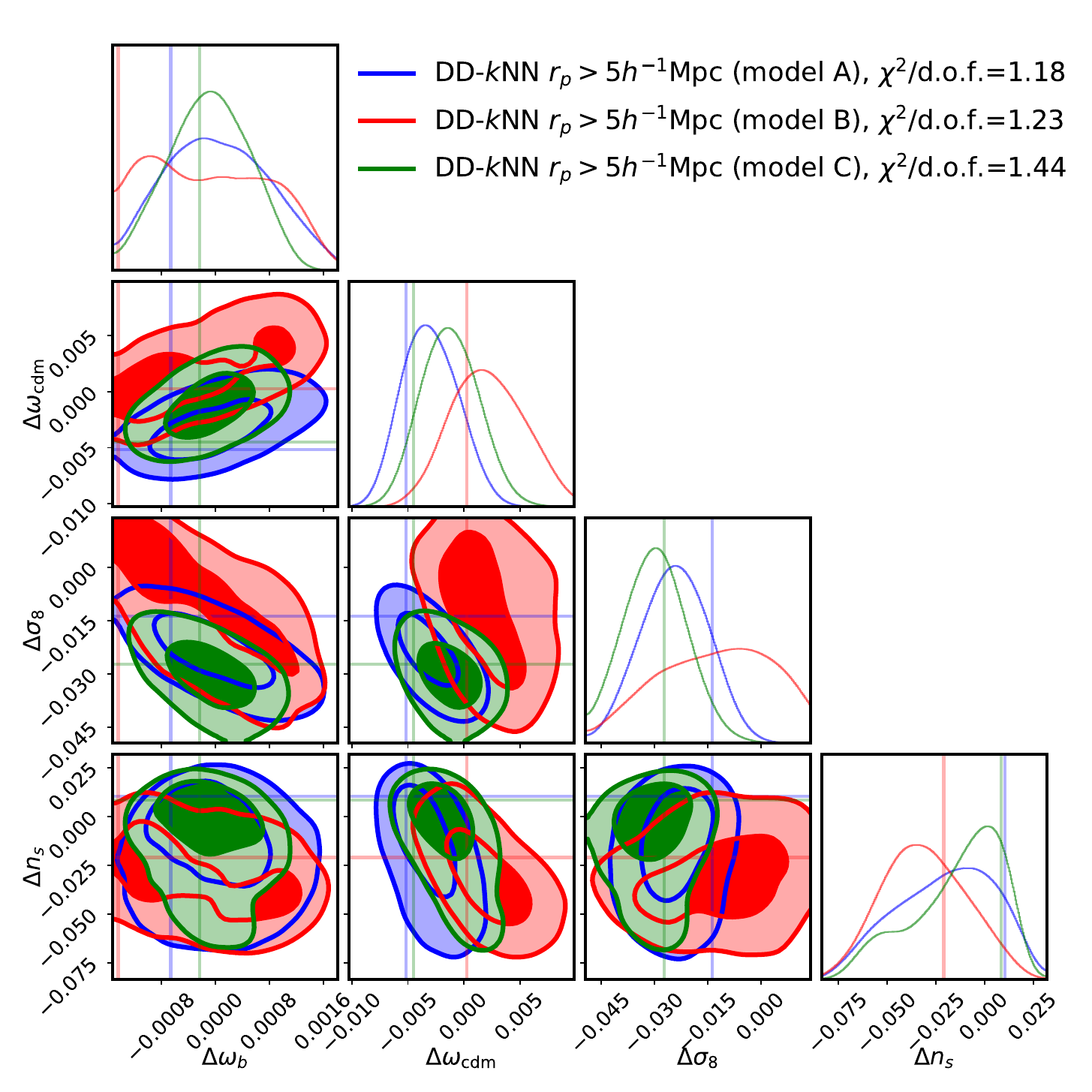}
    \vspace{-0.3cm}
    \caption{The $\Lambda$CDM cosmology posteriors inferred with the DD-$k$NN data vector at $r_p > 5h^{-1}$Mpc using three different HOD models. Again, we use $k = 2,3,...,9$, covering scales $5h^{-1}\mathrm{Mpc} < r_p < 63h^{-1}\mathrm{Mpc}$ and $0.5h^{-1}\mathrm{Mpc} < r_\pi < 31.6h^{-1}\mathrm{Mpc}$. The three contours show good agreement, with significantly improved goodness-of-fit compared to Figure~\ref{fig:corner_ddknn}. \sandy{The contours correspond to $68\%$ and $95\%$ confidence intervals.}}
    \label{fig:corner_ddknn5}
\end{figure*}

Figure~\ref{fig:chi2_scale} shows the best-fit $\chi^2$ as a function of minimum scale cuts. The first two columns show the fiducial cuts used for the RD-$k$NN and DD-$k$NN data vectors. The last two columns show best-fit $\chi^2$ when progressively larger minimum scale cuts are applied to the DD-$k$NNs. We see that we only achieve a good fit $\chi^2$/d.o.f. $\approx 1$ at $r_p > 5h^{-1}$Mpc. 
Figure~\ref{fig:corner_ddknn5} shows the DD-$k$NN cosmology posterior when an additional cut $r_p > 5h^{-1}$Mpc is applied. We see that indeed, the posterior constraints of the three models are now consistent with each other. Thus, we conclude that our current HODs can only reliably model DD-$k$NN at scales greater than $5h^{-1}$Mpc, and we only report our DD-$k$NN constraints at $r_p > 5h^{-1}$Mpc for the blind challenge. However, we still include the full-scale constraints in our discussions to motivate future work. 

The fact that the current HOD models failed to fit DD-$k$NNs on very small scales suggests that the DD-$k$NNs are highly informative on HOD modeling. In \cite{2023Yuan}, we demonstrated that the DD-$k$NN is significantly more constraining on HOD and assembly bias than the standard redshift-space 2PCF. We showed that the 2PCF is a strict marginalization of the DD-$k$NN. Conceptually, DD-$k$NN captures extra information because while the 2PCF encodes the average number of neighbors of a galaxy at a certain distance, the DD-$k$NN additionally encodes the $k$ ordering of neighbors at any distance. Thus, DD-$k$NN additionally captures the phase space and topological configuration information. 

There are several potential reasons why our model cannot produce the DD-$k$NN below $5h^{-1}$Mpc. First of all, without additional information on the galaxy sample that would otherwise be available in a realistic survey, we cannot meaningfully improve our galaxy--halo modeling. For example, the photometric selection allows us to assess the completeness of the sample at different magnitudes. The colours and morphology would inform us of the galaxy type. The spectra can additionally give us statistical descriptions of key galaxy properties such as stellar mass, star formation rate, and metallicity. All these sources of information are useful in building up priors on the galaxy model. For example, \cite{2022Wang} showed that a magnitude-limited sample in SDSS can be robustly described by an HOD model that includes assembly bias. A series of recent studies have also combined realistic selections with state-of-the-art hydrodynamical simulations and semi-analytic models of galaxy formation to build up priors on the appropriate galaxy--halo connection model \citep[e.g.][]{2021Xu, 2022mHadzhiyska, 2022m2Hadzhiyska, 2022Yuan}. 

An additional systematic effect is that the target mock is generated with a different simulation code and different halo finder, which can all produce small-scale features that the DD-$k$NN is potentially sensitive to. This is an issue that we would need to be able to marginalise over in our forward model if we want to exploit 1-halo scales. We propose to assess this point in a future post-unblinding analysis where we can disentangle our uncertainty about the true HOD and the systematic effects associated with simulations and halo finding. 

To summarise, the DD-$k$NNs are potentially highly sensitive to small-scale effects such as the details of galaxy--halo connection modeling, halo finding, and simulation codes. For the sake of this analysis, we can protect ourselves from these systematics by applying a relatively conservative scale cut of $r_p > 5h^{-1}$Mpc. In contrast, the RD-$k$NNs appear to be significantly less sensitive to these effects as long as we choose appropriate query spacing. 

\begin{figure*}
    \hspace{-0.3cm}
    \includegraphics[width=1.0\textwidth]{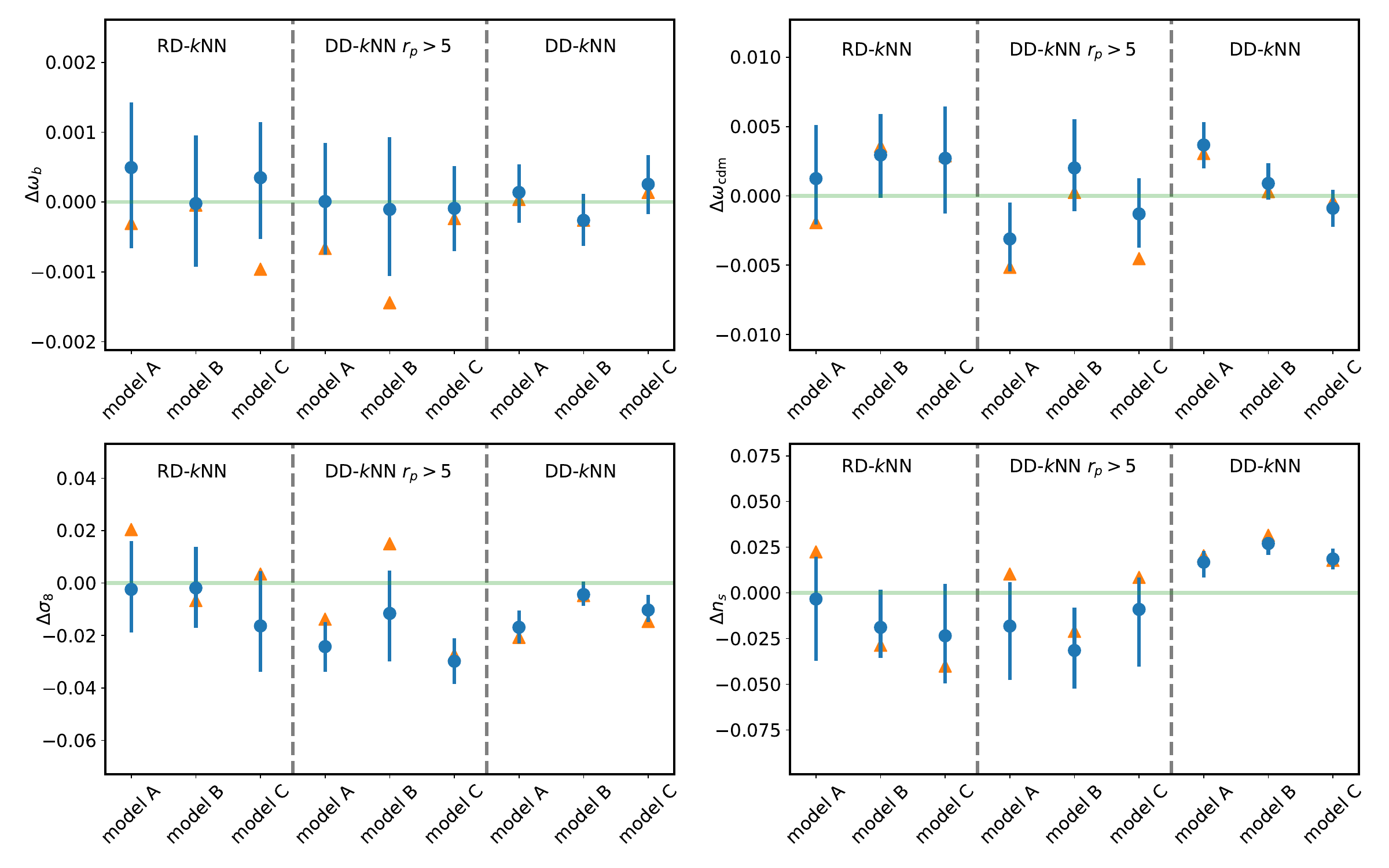}
    \vspace{-0.3cm}
    \caption{The marginalised $\Lambda$CDM cosmology posteriors inferred with the two $k$NN data vectors and using three different HOD models. The blue error bars indicate the (16, 50, 84)$\%$ posterior constraints, and orange triangles show the maximum likelihood point. The $x$-axis iterates through the HOD models, with the fits vertically grouped by data vector. The parameter values are blinded. There is excellent agreement between the RD-$k$NN and DD-$k$NN ($r_p > 5h^{-1}$Mpc). The full-scale DD-$k$NN ($r_p > 0.67h^{-1}$Mpc) also yields reasonably consistent constraints, albeit with significantly tighter error bars, and moderately biased results in some parameters.}
    \label{fig:1d_cosmo}
\end{figure*}

Figure~\ref{fig:1d_cosmo} summarises the cosmology constraints from the different analysis choices. These include all blind analyses we carried out for the challenge. The $y$-axis reports the difference between the inferred and true values, with the green lines denoting 0. The blue lines show the posterior mean and $1\sigma$ constraints. The orange triangles show the maximum likelihood points. Immediately, all our results are remarkably consistent with each other. We largely derive unbiased cosmology constraints, except in $n_s$ when we use the full-scale range in DD-$k$NN. The occasional disagreement between the maximum likelihood points and the posterior constraints indicates potential prior volume / projection effects. This warrants more careful investigation in future application of our methods. 

Finally, we highlight the stringent constraints on $\sigma_8$ from both $k$NN statistics. The full DD-$k$NN constrains $\sigma_8$ error bars to less than 0.01. Even after the scale cuts, the $1\sigma$ constraints are still below $0.02$ for both RD-$k$NN and DD-$k$NN, significantly stronger than existing $\sim 5\%$ constraints from galaxy clustering in BOSS/SDSS \citep[e.g.][]{2017Alam, 2021Lange, 2021Kobayashi, 2022bYuan, 2022Zhai}. While this extra precision comes directly from the increased volume in this mock challenge relative to BOSS, what we have demonstrated is that we can effectively take advantage of the extra volume on non-linear scales, which 
is essential for taking full advantage of the data that will be imminently available with DESI. These constraints also demonstrate the ability of our novel statistics to capture small-scale information. Assuming we are able to construct more informed galaxy--halo connection models with DESI, this should enable us to push our analysis down to smaller scales and is expected to yield even tighter constraints. 

\subsection{Validation of galaxy--halo connection modeling}
\label{subsec:hodval}

To derive reliable cosmology constraints, it is essential to demonstrate that the galaxy--halo connection model is sufficiently flexible to describe the relevant features in the galaxy--halo physics, but not so flexible that it dilutes the cosmology constraints. There are several ways one can inform and validate the galaxy--halo connection model. We summarise them here:
\begin{itemize}
    \item Simulated models: Simulated galaxy models such as hydrodynamical simulations and semi-analytic models are excellent sandboxes to gain physical intuitions about the galaxy sample and identify necessary ingredients in their galaxy--halo connection models. \sandy{However, hydrodynamical simulations are currently too small to calibrate our models to the precision necessary for next-generation surveys. Semi-analytical models are cheaper and can be generalised to larger volumes, but they are still too expensive right now to fairly sample the model space while maintaining cosmological volumes (see \citealt{2023Perez} for the current state-of-the-art).} 
    \item Galaxy observations: Leveraging existing observations and information about the galaxy sample is a powerful and data-driven way to understand and constrain the galaxy--halo connection. For example, one can directly constrain the halo mass given observed galaxies via gravitational lensing. There are also patches of the sky that are observed by a variety of facilities (such as the COSMOS field), yielding deep images across a wide wavelength domain and spectra. One can leverage these datasets to understand the completeness, mass function, and other galaxy properties. These observations can be richly informative but require careful modeling to connect to theory models of dark matter. 
    \item Mock challenges: Showing unbiased cosmology constraints on simulated galaxy mocks is perhaps the most direct way to show robustness. This was the primary motivation for the Beyond-2p challenge. However, the mocks are still generated relying on some assumptions about galaxy formation physics or the galaxy--halo connection. Thus, while these tests help develop confidence (the larger the variety of galaxy models, the better), it does not rule out the possibility that unforeseen galaxy features in the real Universe can bias the inference.
    \item Statistical validations: Statistical tests such as goodness-of-fit and cross-validation are model-agnostic ways to show consistency with data. The idea is to confront the model with all relevant aspects of the data and show that the model can describe the data without internal tensions. \sandy{In this section, we will first test the flexibility of our HOD models with goodness-of-fit metrics. Then we select the ``correct'' HOD model via cross-validations, where we break the data into multiple parts that contain different information and demonstrate that the model calibrated on one part can consistently predict another part. A sufficiently flexible model with correct physical assumptions should achieve good fits and self-consistently predict all relevant aspects of data. In the current work, because we have access to multiple summary statistics that access different subsets of the information content of the galaxy field, we can conduct this cross-validation test by fitting the model on one summary statistic and checking if the constrained model can predict another statistic.}
    
\end{itemize}

In the context of this mock challenge, we cannot leverage simulated galaxy models or galaxy observations to learn the galaxy--halo connection model, but we can conduct statistical tests to assess model validity. 
To demonstrate this point, we employed three different HOD models in our analysis, adding different combinations of assembly bias and baryon feedback prescriptions to the vanilla HOD. The objective of this section is to compare the three models with goodness-of-fit metrics and cross-validation tests to show that one of the three models is preferred by the data and dependable on the relevant scales. 

First, we compare the three models on the basis of their $\chi^2$/d.o.f. (see Figure~\ref{fig:chi2_scale} and the legends of Figure~\ref{fig:corner_rdknn}--\ref{fig:corner_ddknn5}). For RD-$k$NN, all three models achieve good fits to the mock data, with models A and B showing slightly better $\chi^2$. For DD-$k$NN, models B and C yield significantly better fits on small scales. With the $r_p > 5h^{-1}$Mpc cut, the models A and B result in better fits. These comparisons are not conclusive but show that model B performs consistently well in configurations. We do not show joint fits of the DD-$k$NN and RD-$k$NN due to limitations in modeling the joint covariance matrix. 

The reduced $\chi^2$ metric gives a rough characterisation of the goodness-of-fit but offers a poor assessment of possible over-fitting. Instead, we perform cross-validation tests where we take the fits on one summary statistic to predict another summary statistic, relying on the notion that the model with the right physical assumptions should extrapolate to predict additional data. Our summary statistics are particularly well suited for this test because the different summary statistics capture disjoint information in the density field. Specifically, while the DD-$k$NN and the 2PCF capture the clustering information and emphasise the structure of density peaks on small scales, the RD-$k$NN capture the density information and transition between voids and density peaks \citep{2023Yuan}. It is thus particularly useful to conduct cross-validations between a clustering statistic such as (DD-$k$NN or the 2PCF) and a density statistic (such as RD-$k$NN). \sandy{This idea is not completely new. A similar cross-validation was done in \cite{2022Wang} between clustering and counts-in-cell statistics to validate galaxy assembly bias models. A similar cross-validation scheme was also used in recent assessments of potential tensions between galaxy clustering and galaxy-galaxy lensing statistics \citep[e.g.][]{2017Leauthaud, 2019Yuan, 2020Lange, 2023Contreras}.}

\begin{figure*}
    \hspace{-0.3cm}
    \includegraphics[width=1.0\textwidth]{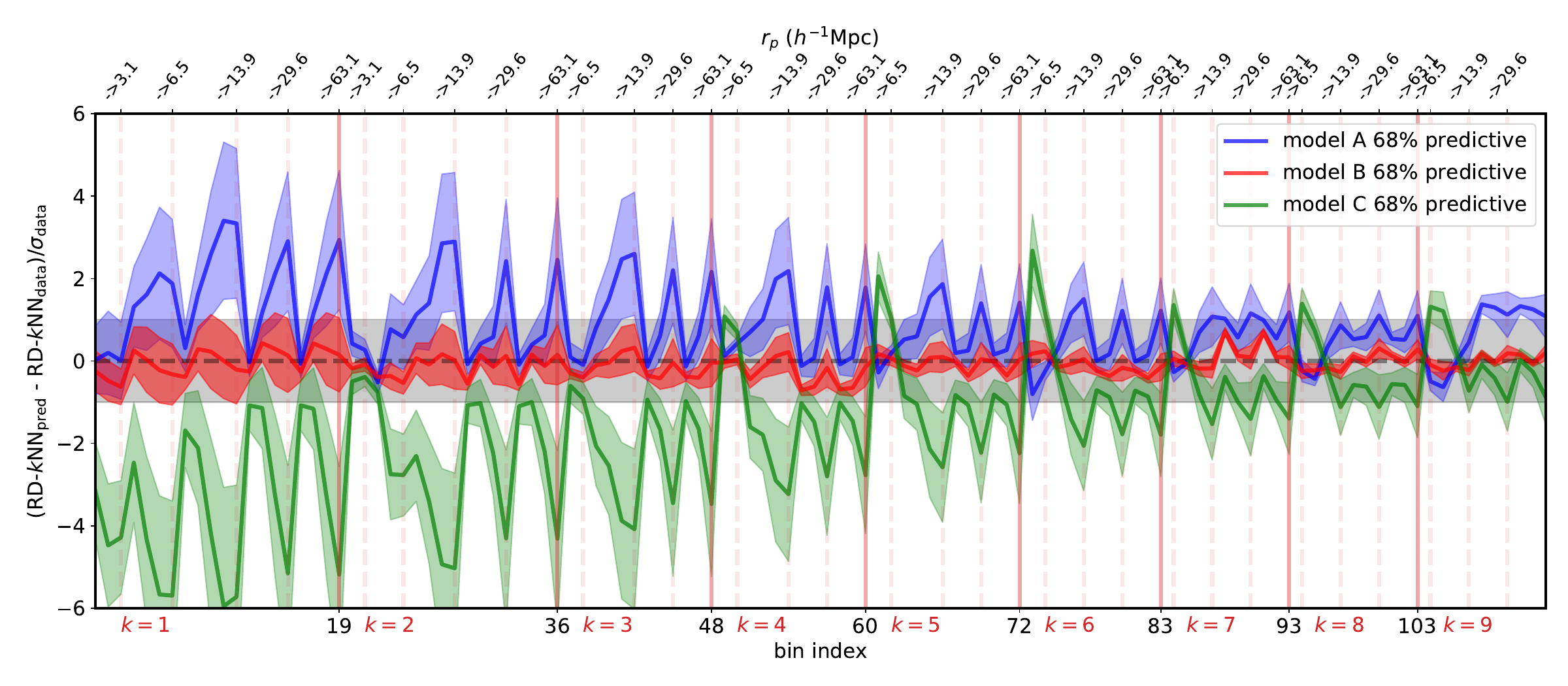}
    \vspace{-0.3cm}
    \caption{The predictive distribution of the RD-$k$NNs when the models are fitted to the DD-$k$NNs. The $y$-axis shows the difference between the RD-$k$NN prediction and the true RD-$k$NN on the blind mock, normalised by the diagonal jackknife errors. The three shaded lines show the $1\sigma$ predictive distributions of the three different HOD models. The $x$-axis shows the bin indices in RD-$k$NN. Each large block separated the solid vertical dashed lines correspond to different $k$s. The sub-blocks separated by dashed vertical lines denote different $r_p$s; each sub-block cycles through the $r_\pi$s. Model B makes a consistent prediction for RD-$k$NNs given DD-$k$NNs, whereas the other two models do not. }
    \label{fig:dd-rd}
\end{figure*}

In Figure~\ref{fig:dd-rd}, we show the RD-$k$NN predictive distribution derived when the models are fit to the DD-$k$NNs. In particular, the $y$-axis shows the difference between the RD-$k$NN prediction with the three models and the true RD-$k$NN on the blind mock, normalised by the jackknife errors. The three shaded lines represent the $1\sigma$ predictions for the three HOD models. Model B successfully predicts the correct RD-$k$NN, while the two other models fail. \sandy{This indicates that model B is both sufficiently flexible to describe the relevant aspects of the data (good $\chi^2$/d.o.f.) and carries the right physical assumptions such that it can predict additional statistics it is not fit on.} It is also worth noting that all three models make reliable predictions at the beginning of each sub-block (shown by vertical dashed lines), corresponding to small $r_\pi$ separation, but fail at larger $r_\pi$ values. Thus, it appears that model B might be better at capturing the correct redshift-space distortion compared to the other models. All three models make consistent predictions at large $k$ (as in $k$NN, not spatial frequency), which makes sense because large $k$ orders probe larger scales and contain less information about the galaxy--halo connection. 

\begin{figure}
    \hspace{-0.3cm}
    \includegraphics[width=0.5\textwidth]{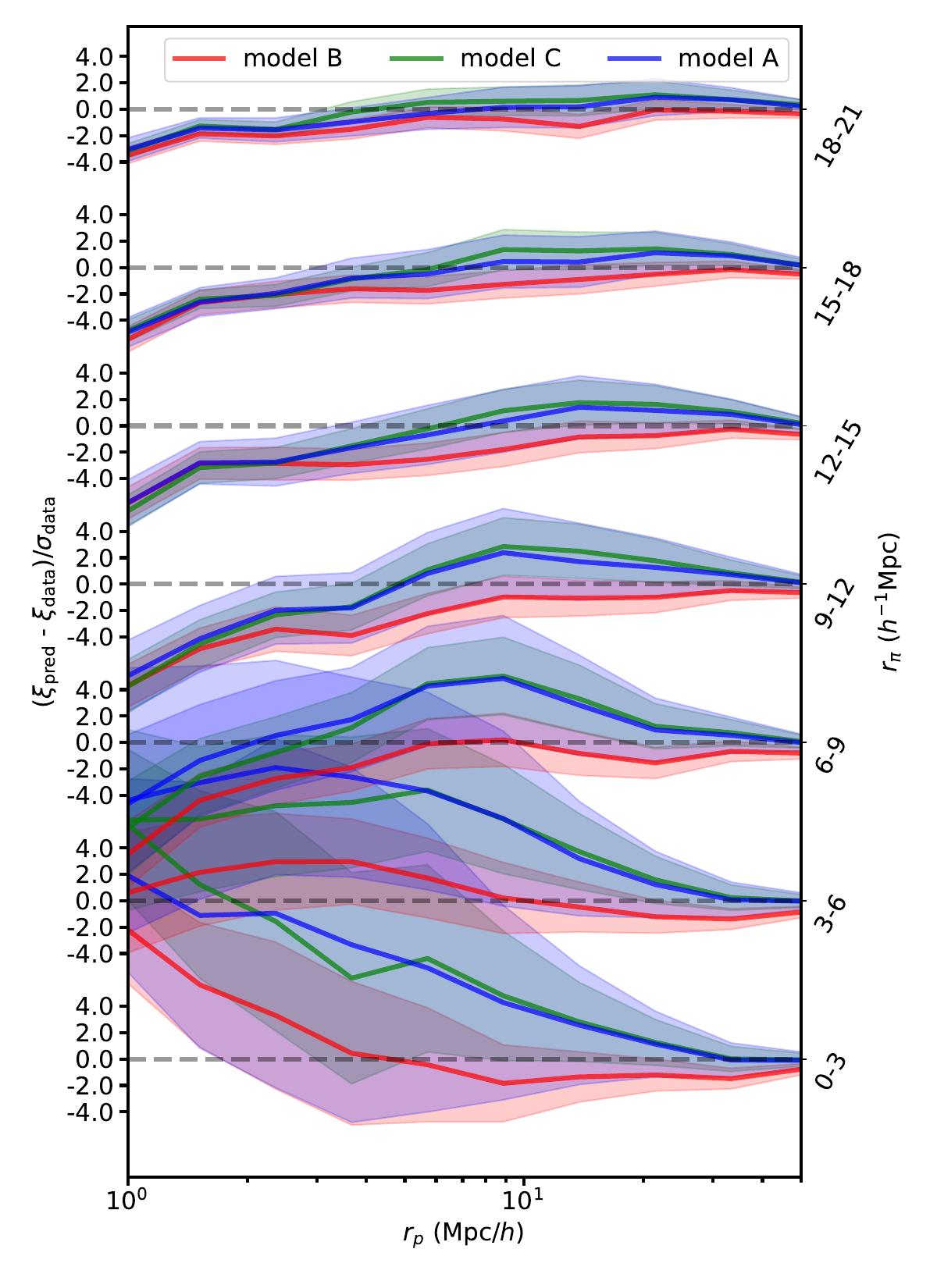}
    \vspace{-0.5cm}
    \caption{Redshift-space 2PCF $\xi(r_p, r_\pi)$ predictive distribution in the case where the models are fitted to the RD-$k$NNs. The $y$-axis on the left side shows the difference between the predicted $\xi(r_p, r_\pi)$ and the true $\xi(r_p, r_\pi)$, normalised by the jackknife errors. We have broken the 2D $\xi(r_p, r_\pi)$ data vector into different rows, each row corresponding to a $r_\pi$, where the $r_\pi$ ranges for each row are shown on the $y$-axis to the right. Model B successfully predicts the 2PCF down to smaller $r_p$ scales than the other two models. }
    \label{fig:rd-tp}
\end{figure}

Figure~\ref{fig:rd-tp} shows the redshift-space 2PCF $\xi(r_p, r_\pi)$ predictive distribution when the models are fitted to the RD-$k$NNs. The $y$-axis on the left side shows the difference between the predicted $\xi(r_p, r_\pi)$ and the true $\xi(r_p, r_\pi)$, normalised by the jackknife errors. We have broken the 2D $\xi(r_p, r_\pi)$ data vector into different rows, each row corresponding to a $r_\pi$. We see that all three models can predict the 2PCF at large $r_p$, but none of the three models correctly predict the 2PCF all the way down to $r_p = 1h^{-1}$Mpc. However, comparing the three models, model B predicts the 2PCF consistently down to $r_p = 2h^{-1}$Mpc whereas the other two models deviate from the true values at larger scales. 

\sandy{To summarise, model B is strongly preferred by the mock data, both because of good $\chi^2$/d.o.f. and its superior ability to predict statistics that it was not fit on. Model B performed better in both the RD-$k$NN prediction and the 2PCF prediction. It was particularly telling when we fitted the three different HOD models on DD-$k$NN and only model B gave a fully consistent prediction of the RD-$k$NN. }

The trend in the disagreement shown in Figure~\ref{fig:rd-tp} also shows that the current failure to model scales smaller than $r_p < 3h^{-1}$Mpc might be due to insufficient flexibility in modeling the small-scale finger-of-god. While we have included velocity bias in both central and satellite galaxies in our velocity model, the parameterisation only allows for a constant fractional shift relative to the underlying dark matter. \sandy{There exists evidence that a more sophisticated velocity bias model is needed. For example, \cite{2017Ye} reports significant mass dependencies in hydrodynamical models.} Additionally, the differences between simulation codes and halo finders can introduce significant complexities in the galaxy velocities. This highlights the need for more flexible velocity modeling in our pipeline. 
\sandy{Several other possible reasons for poor fits at very small scales include unrealistic models used in the mock production and underestimated covariances. It is particularly important to further validate our covariance matrices due to their large size and significant off-diagonal contribution. We reserve that for a future paper. }

 \section{Discussion}
 
 \begin{figure*}
    \hspace{-0.3cm}
    \includegraphics[width=0.83\textwidth]{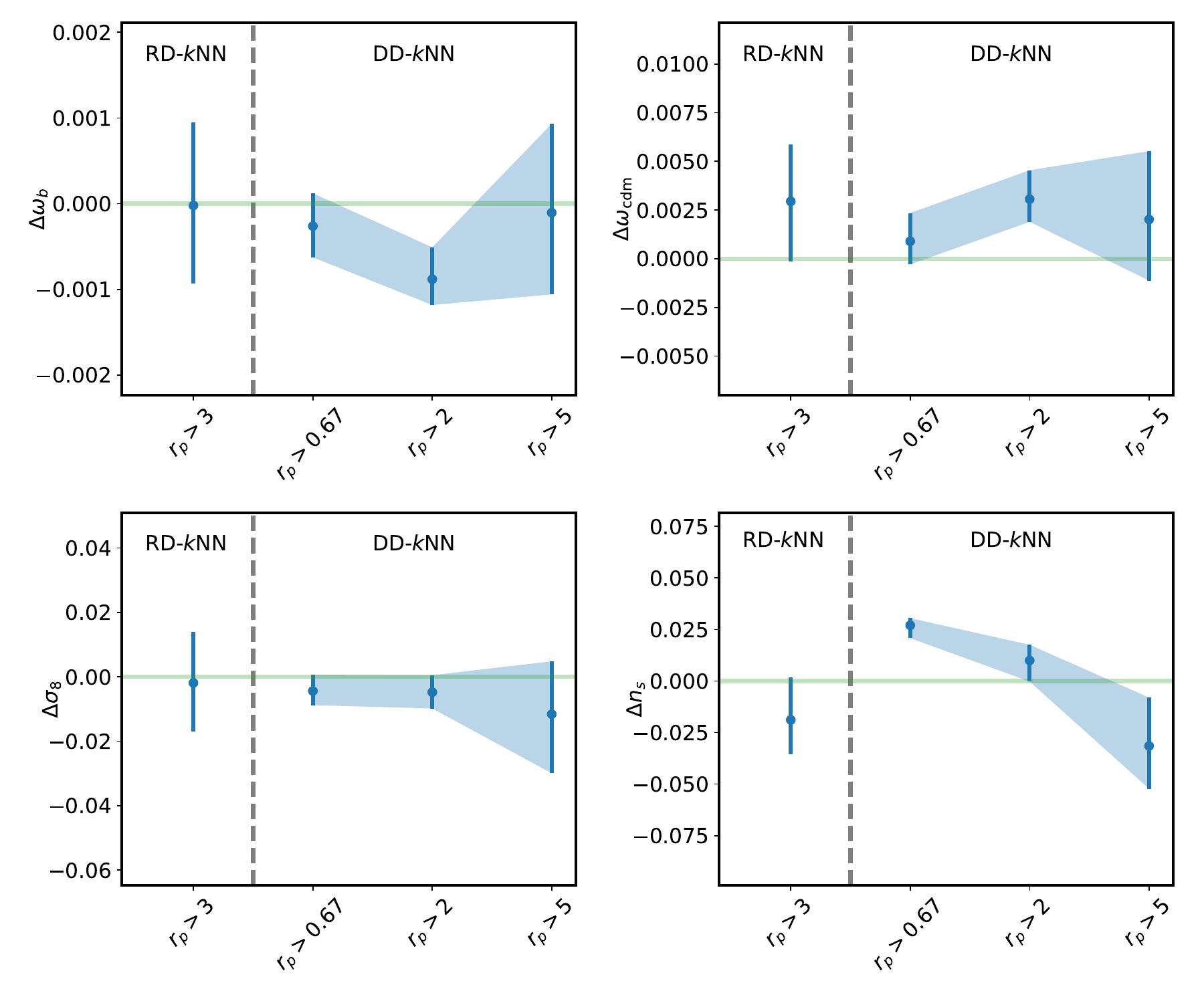}
    \vspace{-0.3cm}
    \caption{The marginalised $\Lambda$CDM cosmology posteriors when using different minimum scale cuts for the DD-$k$NN statistics. For this test, we have fixed the HOD model to model B. The constraining power of DD-$k$NN strongly depends on minimum scale cut. We do not vary the minimum scale of RD-$k$NN as it would require extensive re-computation. }
    \label{fig:1d_cosmo_scale}
\end{figure*}
So far, we have presented the methodology and results of applying our $k$NN analysis pipeline to the Beyond-2p mocks. While we derive strong and unbiased constraints on cosmology, we highlight a few caveats and provide outlook for future work in this section. 

\label{sec:discuss}
    \subsection{Addressed and unaddressed systematics}
The fact that we derive unbiased cosmology from deeply non-linear scales is significant for a couple of reasons. We are among the first to show that a simulation-based method for non-linear scales can recover unbiased cosmology results when confronted with mock data generated with a completely different simulation. This shows that our modeling framework can be resilient against several simulation-related systematics, including gravity codes, simulation resolution, and halo finding \citep[e.g.][]{2008Heitmann, 2011Knebe, 2022Grove}. We attribute this resilience to the careful choice of HOD models and scale cuts. This conclusion is made stronger by the fact that our mock data was blind and the true cosmology was far from Planck, which shows that we are not biased by priors. 

However, one key systematic that this exercise has not fully addressed is galaxy--halo connection modeling. Both the models and the mock data are generated from HOD models. We first note that the HOD itself is actually a broad framework and there is a large amount of potential systematics that can arise from different implementations of HODs \citep[e.g.][]{2016Hearin}. In its most general interpretation, the HOD model simply implies a probabilistic occupation based on halo properties \citep[see][for a review]{2018Wechsler}. Thus, there is a large amount of freedom in terms of the specific function form of the HOD, the probability distribution, and the halo properties used. Moreover, the fact that the mock data and the models used different simulations and halo finders means that the behavior of the same HOD can be different in complex ways. All this is to say that even within the HOD framework, it is non-trivial to recover unbiased cosmologies from two different HOD implementations on top of two different halo finders. 

Nevertheless, the HOD is an empirical model that is necessarily wrong in detail when compared to the real Universe \citep[e.g.][]{2020Hadzhiyska, 2021Xu, 2022Beltz-Mohrmann}. Thus, it is important to repeat this exercise with blind mocks constructed from more sophisticated galaxy--halo connection models. \sandy{One good and available option is to use subhalo abundance matching \citep[SHAM][]{2006Vale, 2017Lehmann}, which relies on the assumed correlation between some galaxy property with some subhalo property to assign galaxies. SHAM is meaningfully different from the HOD as it is based on subhalos instead of halos, and it does not rely on analytic forms. However, SHAM has two drawbacks. First, it is similar to HODs in the sense that it is still commonly based on halo mass or other proxies of mass. Second, it also requires large high-resolution simulations with effective subhalo finding, which can be expensive.} Another option is to use machine learning models to paint galaxies over from hydrodynamical simulations or semi-analytic models \citep[e.g.][]{2021Delgado, 2021Xu, 2022Lovell, 2022McGibbon, 2023Chittenden}. These approaches automatically incorporate a variety of subtleties in galaxy modeling and can be a powerful tool to generate complex and realistic mocks. A third option is to use physically motivated empirical models such as UniverseMachine and Diffstar \citep{2019Behroozi, 2023Alarcon}. These models do not rely on assumptions built into external simulations and provide a self-consistent way of modeling galaxy assembly. Both the second and third options should be highly differentiated from the HOD and would serve well to test against galaxy--halo modeling systematics. 

Finally, this analysis does not address observational systematics such as survey incompleteness due to survey windows and fiber collisions, or evolution effects over finite redshift ranges. These effects are non-trivial to correct for when calculating novel summary statistics such as the $k$NNs. However, in principle we can forward model these effects using simulation lightcones \citep{2022cYuan, 2022Hahn}. Specifically, in \cite{2022cYuan}, we present both a flexible and performant framework to explore HOD and redshift evolution models on lightcones, and techniques to correct for survey boundary and fiber collision effects. Similar realistic lightcone mocks were also constructed and tested for large photometric surveys in \cite{2019DeRoseb}, \cite{2018MacCrann}, and \cite{2021To}. Combining the forward modeling framework and the modeling techniques laid out in this paper will be powerful in exploiting non-linear scale information in upcoming spectroscopic surveys.

 \subsection{The impact of scale cuts}
 \label{subsec:dicsuss_scale}
In section~\ref{subsec:posteriors}, we chose a fiducial scale cut of $r_p > 5h^{-1}$Mpc for the DD-$k$NN statistics. Such a scale cut is indeed necessary for this specific analysis as our galaxy--halo connection model is not sufficiently flexible to self-consistently predict features on smaller scales. However, we also pointed out that in a realistic survey, we would have considerably more information on the target galaxy sample to better inform the galaxy--halo connection model. Thus, it is interesting to discuss how our constraints would improve as we push down to smaller scales with improved models. 

Figure~\ref{fig:1d_cosmo_scale} shows the 1D marginalised constraints on cosmology as we include smaller $r_p$ in DD-$k$NNs. Clearly, the constraining power of the DD-$k$NN statistics depends strongly on the minimum scale cuts. At $r_p > 0.67 h^{-1}$Mpc, we see a 4 times improvement in the $\sigma_8$ and $n_s$ constraints and a 2-3 times improvement in the constraints on the density parameters. This demonstrates that while we can already derive highly competitive constraints with our fiducial scale cut of $r_p > 5 h^{-1}$Mpc, the true power of the DD-$k$NN statistics lies in the smaller scales. This strongly motivates the need for a more informed yet flexible model of galaxy--halo connection in the advent of next-generation surveys such as DESI. We note that these constraints are derived with a fixed galaxy--halo connection model that we already showed to be insufficient to describe the smaller scales. Thus, while the improvement in constraining power is informative of the next priority in further developing simulation-based models, it should not be interpreted as a detailed forecast for future analysis. The exact constraining power will depend on the modeling specifics. 

It is also important to briefly discuss the impact of maximum scale cuts. In this analysis, we are exclusively interested in non-linear scales, so we chose a modest maximum scale cut of $r_p < 63h^{-1}$Mpc, which reaches into the linear regime but does not capture key large-scale observables such as the BAO or the matter-radiation equality scale. In principle, we can gain sensitivity to those scales by extending the $r_p$ range and including much higher $k$ orders. We expect this expansion to significantly improve our constraints on the mass density parameters but also to have minimal effect on our $\sigma_8$ and $n_s$ constraints, as those come predominantly from small scales and RSD. We reserve a full-scale $k$NN analysis for a future paper.

\section{Conclusions}
\label{sec:conclude}
The key to unlocking the cosmological information on non-linear scales lies in employing simulation-based models and high-order summary statistics. 
In this paper, we present our methodology for recovering robust cosmology constraints from non-linear scales, leveraging two flavors of the $k$NN statistics, one capturing the density field (RD-$k$NN) and the other capturing the clustering (DD-$k$NN). We confront a blind mock with three different HOD models, including varied prescriptions of assembly bias and baryonic effects built on top of the \textsc{AbacusSummit} cosmology suite. We demonstrate several key points in this paper:
\begin{itemize}
    \item This is the first time a simulation-based model has been shown to derive strong and unbiased cosmology constraints in a blind mock challenge of comparable volume to a modern spectroscopic survey. The power of this result is strengthened by the fact that the mock uses different simulation and halo codes than the input data, and by the fact that the underlying cosmology is significantly different than Planck. 
    
    \item We demonstrated the flexibility and resilience of our approach using multiple HOD models by employing a mix of goodness-of-fit metrics and cross-validation tests. These tests give a clear preference for one of the models, which is able to reproduce a second statistic after being fit to a first. On real data, more information may be available to further inform the HOD model tests.
    \item We use goodness-of-fit metrics to establish minimum scale cuts for our analysis. The RD-$k$NN statistic emerged as less prone to non-linear scale modeling errors. However, the DD-$k$NN proved to be considerably more constraining when we employ improved models of small scales.
\end{itemize}

This work underscores the potential of using beyond-2p statistics on non-linear scales for cosmology, especially in the context of the next generation of spectroscopic surveys. We argue that integrating direct insights on the galaxy--halo connection from data and simulations would likely enable analyses on even smaller scales. 

Moving forward, it would be valuable to re-do this analysis with a variety of more sophisticated galaxy--halo connection models that are not based on HODs. That would more convincingly demonstrate the credibility of simulation-based modeling approaches for cosmology. It is also important to test against observational systematics such as survey incompleteness and redshift evolution, for example, using a lightcone-based forward modeling approach like in \cite{2022cYuan} and applying these systematics directly. One can also resort to realistic non-HOD mocks such as in \cite{2019DeRoseb}, \cite{2018MacCrann}, and \cite{2021To}. We look forward to applications of such approaches in future analyses. 

\section*{Acknowledgements}

We would like to thank Elisabeth Krause, Yosuke Kobayashi, Andres Salcedo, Enrique Paillas, and others for useful feedback and suggestions in various stages of this analysis.

This work was supported by the U.S. Department of Energy through grant DE-SC0013718 and under DE-AC02-76SF00515 to SLAC National Accelerator Laboratory, and by the Kavli Institute for Particle Astrophysics and Cosmology. This work was performed in part at the Aspen Center for Physics, which is supported by National Science Foundation grant PHY-2210452.

This work used resources of the National Energy Research Scientific Computing Center (NERSC), a U.S. Department of Energy Office of Science User Facility located at Lawrence Berkeley National Laboratory, operated under Contract No. DE-AC02-05CH11231.

The {\sc AbacusSummit} simulations were conducted at the Oak Ridge Leadership Computing Facility, which is a DOE Office of Science User Facility supported under Contract DE-AC05-00OR22725, through support from projects AST135 and AST145, the latter through the Department of Energy ALCC program.

\section*{Data Availability}

The simulation data are available at \url{https://abacussummit.readthedocs.io/en/latest/}. The \ahod\ code package is publicly available as a part of the \textsc{abacusutils} package at \url{https://github.com/abacusorg/abacusutils}. Example usage can be found at \url{https://abacusutils.readthedocs.io/en/latest/hod.html}.



\bibliographystyle{mnras}
\bibliography{biblio} 




\appendix

\section{Covariance matrices}
\begin{figure*}
    \centering
    \hspace*{-0.5cm}
    \includegraphics[width = 5in]{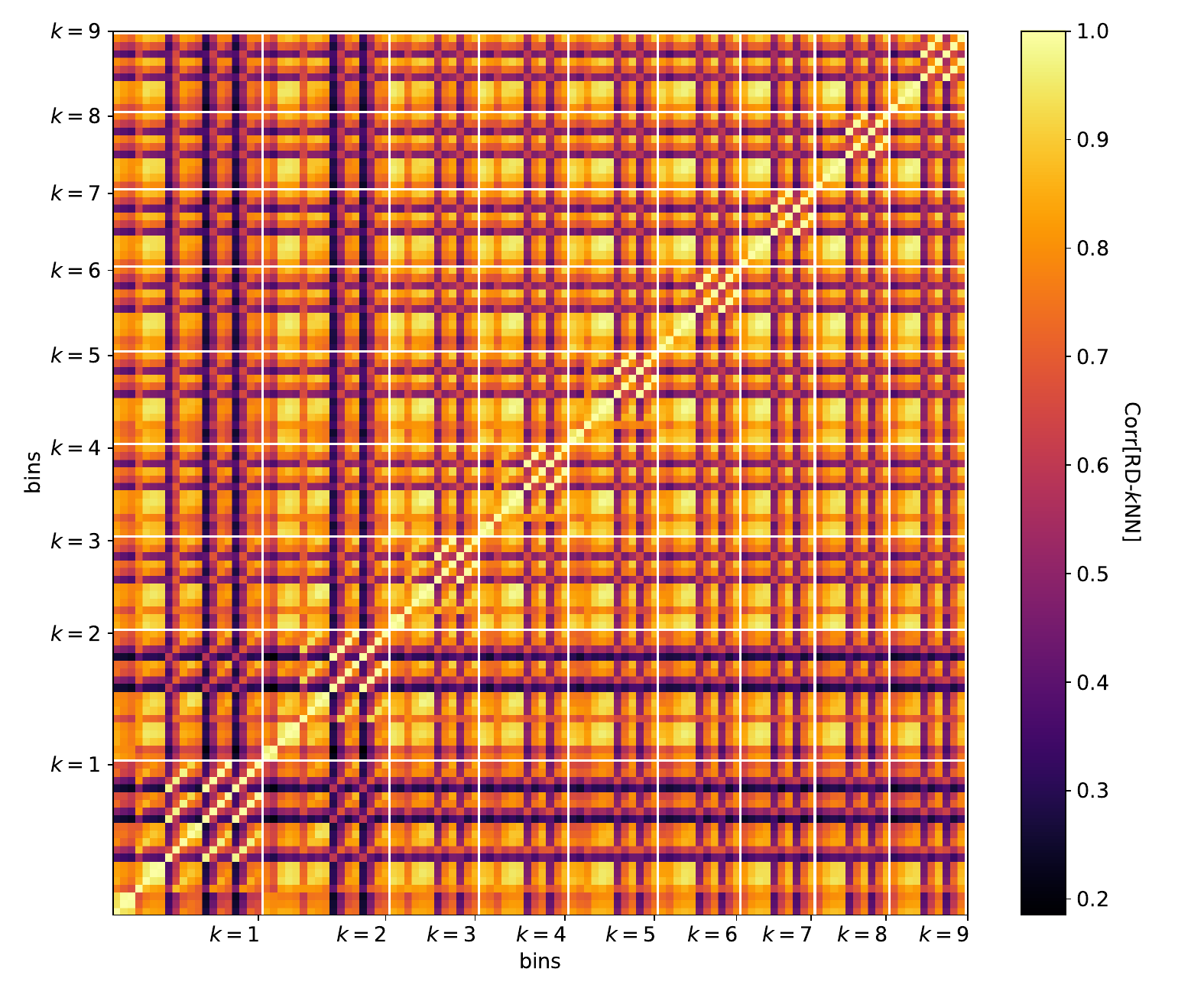}
    \vspace{-0.5cm}
    \caption{Visualizations of the RD-$k$NN correlation matrix. The covariance matrix is computed from 1250 jackknife regions. The ``bins'' represents the flattened indices of $k$NN bins along $k$, $r_p$, and $r_\pi$ axes, in orders of outer to inner cycles. For ease of visualization, we separate the bins into blocks via the vertical and horizontal white lines, where each block corresponds to a $k$. Within each $k$ block, the increasing bin number cycles through $r_\pi$ values at each fixed $r_p$. 
    }
    \label{fig:knns_cov_rd}
\end{figure*}

\begin{figure*}
    \centering
    \hspace*{-0.5cm}
    \includegraphics[width = 5in]{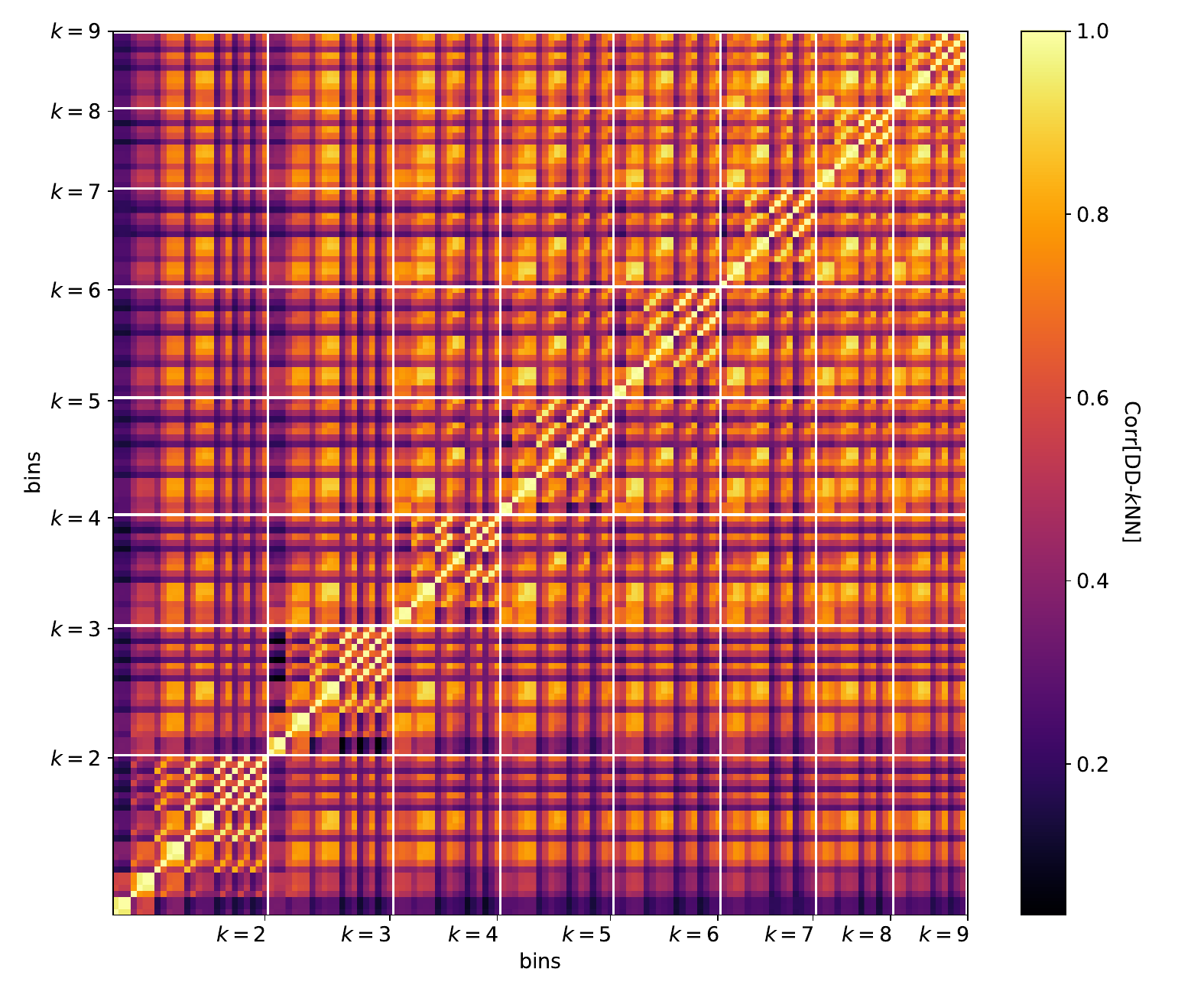}
    \vspace{-0.5cm}
    \caption{Visualizations of the DD-$k$NN correlation matrix. The covariance matrix is computed from 1250 jackknife regions. The indices are arranged in the same way as Figure~\ref{fig:knns_cov_rd}. Note that for DD-$k$NN, the lowest meaningful value is $k = 2$, because the first neighbor is the galaxy itself.}
    \label{fig:knns_cov_dd}
\end{figure*}

Figure~\ref{fig:knns_cov_rd} and Figure~\ref{fig:knns_cov_dd} show the covariance matrices of the RD-$k$NN and DD-$k$NN data vectors calculated from 1250 jackknife regions on the Beyond-2p redshift-space mock. 

\section{Emulator performance}

\begin{figure}
    \hspace{-0.4cm}
    \includegraphics[width=0.55\textwidth]{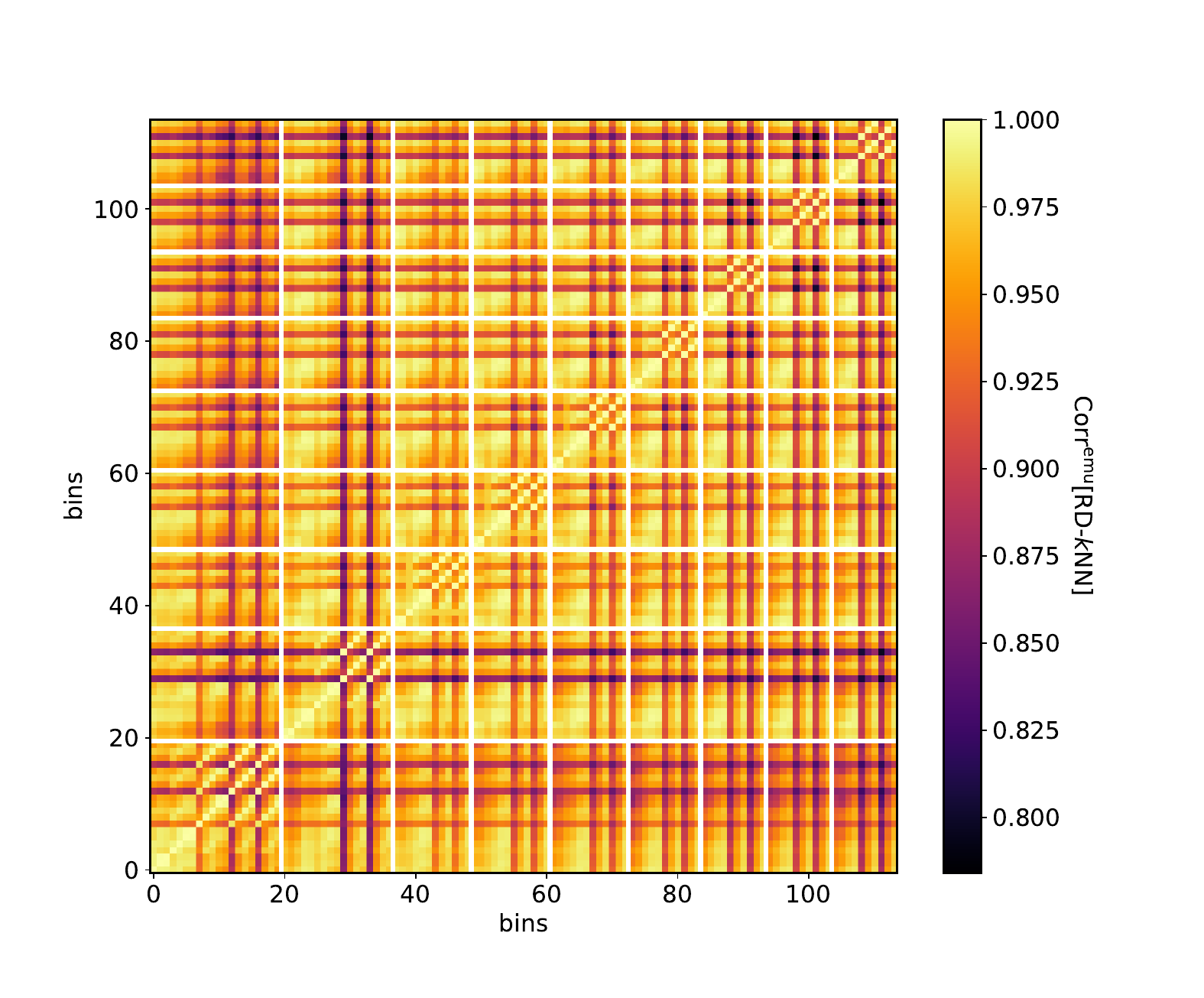}
    \vspace{-0.6cm}
    \caption{The correlation matrix of the RD-$k$NN emulator covariance matrix. This plot is constructed similarly to Figure~\ref{fig:knns_cov_rd}, where we use a white grid to denote different $k$s. The bin indices also assume the same order. }
    \label{fig:corr_rd_emu}
\end{figure}

\begin{figure}
    \hspace{-0.4cm}
    \includegraphics[width=0.55\textwidth]{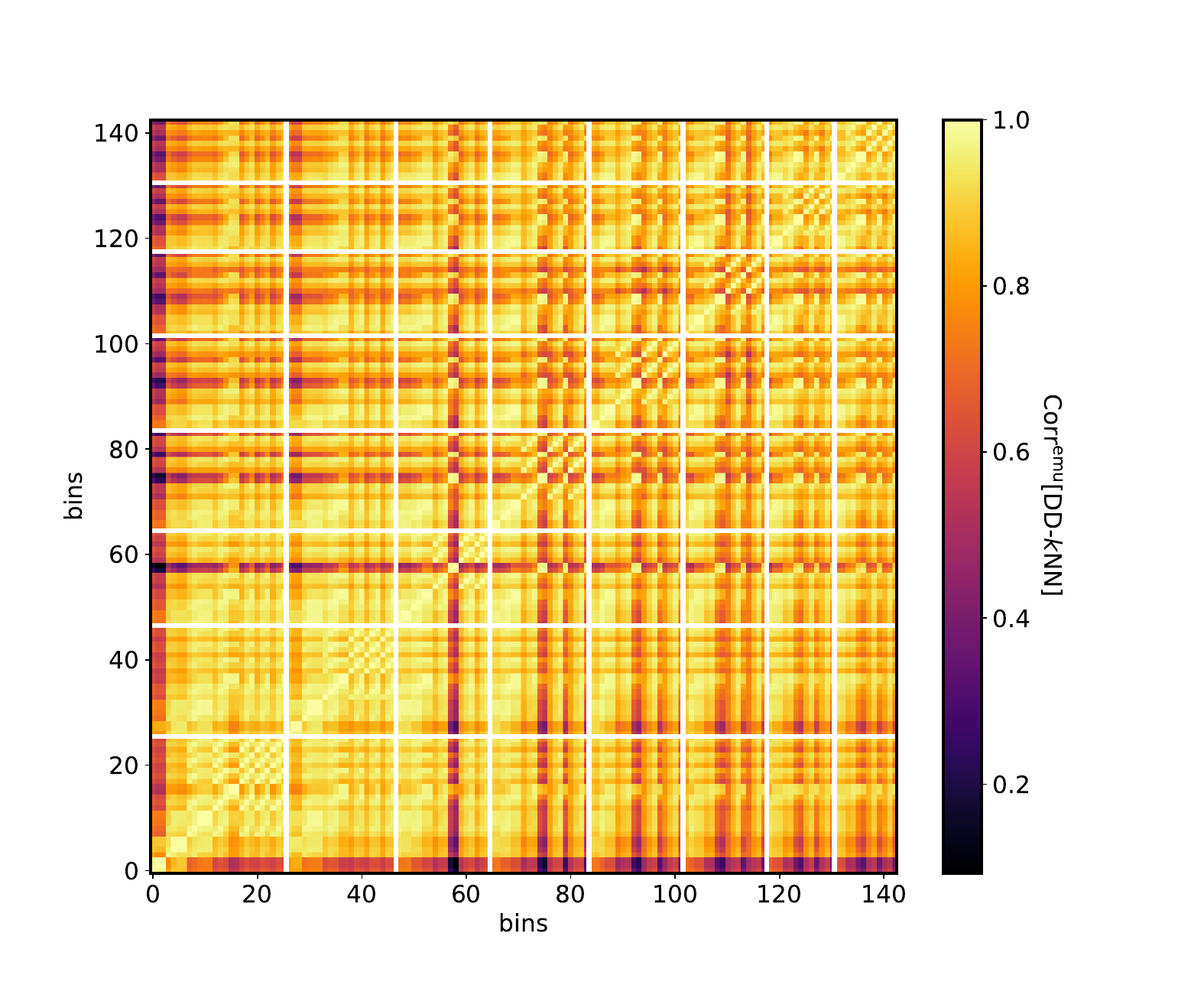}
    \vspace{-0.6cm}
    \caption{The correlation matrix of the DD-$k$NN emulator covariance matrix.}
    \label{fig:corr_dd_emu}
\end{figure}

In this section, we present metrics indicating the performance of the $k$NN emulators. Figure~\ref{fig:corr_rd_emu} and Figure~\ref{fig:corr_dd_emu} show the correlation matrices corresponding to the RD-$k$NN and DD-$k$NN emulator errors. Contrary to the sample variance covariance matrices shown in Figure~\ref{fig:knns_cov_rd} and Figure~\ref{fig:knns_cov_dd}, the emulator errors exhibit larger off-diagonal contributions. The mean absolute error of the emulators is approximately 0.3 and 0.6 that of the sample variance for RD-$k$NN and DD-$k$NN, respectively. That translates to a $10$-40$\%$ increase to the diagonal amplitude of the covariance matrices. 

\bsp	
\label{lastpage}
\end{document}